\documentclass[prb,preprint,amsmath,amssymb,superscriptaddress,notitlepage]{revtex4-1}
\usepackage{epsfig}
\usepackage{graphicx}
\usepackage{color}
\usepackage{bm}
\usepackage{fixme}
\usepackage[colorinlistoftodos,textwidth=2.25cm]{todonotes} 
\usepackage{color,soul}

\begin{document}

\title{Magneto-Seebeck 
microscopy of domain switching in collinear antiferromagnet CuMnAs}
\author{T.~Janda}
\affiliation{Faculty of Mathematics and Physics, Charles University, Ke Karlovu 3, 121 16 Prague 2, Czech Republic}
\affiliation{Institute of Experimental and Applied Physics, University of  Regensburg, Universit\"atsstra{\ss}e 31, 93051 Regensburg}
\author{J.~Godinho}
\affiliation{Institute of Physics, Czech Academy of Sciences, Cukrovarnick\'a 10, 162 00, Praha 6, Czech Republic}
\affiliation{Faculty of Mathematics and Physics, Charles University, Ke Karlovu 3, 121 16 Prague 2, Czech Republic}
\author{T.~Ostatnicky}
\affiliation{Faculty of Mathematics and Physics, Charles University, Ke Karlovu 3, 121 16 Prague 2, Czech Republic}
\author{E.~Pfitzner}
\affiliation{Department of Physics, Freie Universit\"at Berlin, 14195 Berlin, Germany}
\author{G.~Ulrich}
\affiliation{Physikalisch-Technische Bundesanstalt, 38116 Braunschweig and 10587 Berlin, Germany}
\author{A.~Hoehl}
\affiliation{Physikalisch-Technische Bundesanstalt, 38116 Braunschweig and 10587 Berlin, Germany}
\author{S.~Reimers}
\affiliation{School of Physics and Astronomy, University of Nottingham, Nottingham NG7 2RD, United Kingdom}
\author{Z.~\v{S}ob\'a\v{n}} 
\affiliation{Institute of Physics, Czech Academy of Sciences, Cukrovarnick\'a 10, 162 00, Praha 6, Czech Republic}
\author{T.~Metzger} 
\affiliation{Institute of Experimental and Applied Physics, University of  Regensburg, Universit\"atsstra{\ss}e 31, 93051 Regensburg}
\author{H.~Reichlov\'a}
\affiliation{Institut fuer Festkoerper- und Materialphysik, Technische Universit\"at Dresden, 01062 Dresden, Germany}
\author{V.~Nov\'ak} 
\affiliation{Institute of Physics, Czech Academy of Sciences, Cukrovarnick\'a 10, 162 00, Praha 6, Czech Republic}
\author{R.~P.~Campion}
\affiliation{School of Physics and Astronomy, University of Nottingham, Nottingham NG7 2RD, United Kingdom}
\author{J.~Heberle}
\affiliation{Department of Physics, Freie Universit\"at Berlin, 14195 Berlin, Germany}
\author{P.~Wadley}
\affiliation{School of Physics and Astronomy, University of Nottingham, Nottingham NG7 2RD, United Kingdom}
\author{K.~W.~Edmonds}
\affiliation{School of Physics and Astronomy, University of Nottingham, Nottingham NG7 2RD, United Kingdom}
\author{O.~J.~Amin}
\affiliation{School of Physics and Astronomy, University of Nottingham, Nottingham NG7 2RD, United Kingdom}
\author{J. ~S. ~Chauhan}
\affiliation{School of Physics and Astronomy, University of Nottingham, Nottingham NG7 2RD, United Kingdom}
\author{S.~S.~Dhesi}
\affiliation{Diamond Light Source, Chilton, Didcot, United Kingdom}
\author{F.~Maccherozzi}
\affiliation{Diamond Light Source, Chilton, Didcot, United Kingdom}
\author{R.~M.~Otxoa}
\affiliation{Hitachi Cambridge Laboratory, Cambridge CB3 0HE, United Kingdom}
\affiliation{Donostia International Physics Center, 20018 San Sebasti\'an, Spain}
\author{P.~E.~Roy}
\affiliation{Hitachi Cambridge Laboratory, Cambridge CB3 0HE, United Kingdom}
\author{K.~Olejn\'ik} 
\affiliation{Institute of Physics, Czech Academy of Sciences, Cukrovarnick\'a 10, 162 00, Praha 6, Czech Republic} 
\author{P.~N\v{e}mec}
\affiliation{Faculty of Mathematics and Physics, Charles University, Ke Karlovu 3, 121 16 Prague 2, Czech Republic}
\author{T.~Jungwirth}
\affiliation{Institute of Physics, Czech Academy of Sciences, Cukrovarnick\'a 10, 162 00, Praha 6, Czech Republic} 
\affiliation{School of Physics and Astronomy, University of Nottingham, Nottingham NG7 2RD, United Kingdom}
\author{B.~Kaestner}
\affiliation{Physikalisch-Technische Bundesanstalt, 38116 Braunschweig and 10587 Berlin, Germany}
\author{J.~Wunderlich}
\affiliation{Institute of Experimental and Applied Physics, University of  Regensburg, Universit\"atsstra{\ss}e 31, 93051 Regensburg}
\affiliation{Institute of Physics, Czech Academy of Sciences, Cukrovarnick\'a 10, 162 00, Praha 6, Czech Republic}

\maketitle

{\bf 
Antiferromagnets offer spintronic device characteristics unparalleled in ferromagnets owing to their lack of stray fields, THz spin dynamics, and rich materials landscape. Microscopic imaging of antiferromagnetic domains is one of the key prerequisites for understanding  physical principles of the device operation. However, adapting common magnetometry techniques to the dipolar-field-free  antiferromagnets has been a major challenge. Here we demonstrate in a collinear antiferromagnet a thermoelectric detection method by combining the magneto-Seebeck effect with local heat gradients generated by scanning far-field or near-field techniques. In a 20~nm epilayer of uniaxial CuMnAs we observe reversible 180$^\circ$ switching of the N\'eel vector via domain wall displacement, controlled by the polarity of the current pulses.  We also image polarity-dependent 90$^\circ$ switching of the N\'eel vector in a thicker biaxial film, and domain shattering induced at higher pulse amplitudes. The antiferromagnetic domain maps obtained by our laboratory technique are compared to measurements by the established synchrotron-based technique of x-ray photoemission electron microscopy using X-ray magnetic linear dichroism.
}

\bigskip

Writing and reading by electrical and optical means, high speed operation combined with neuromorphic memory characteristics, and novel topological phenomena are among the topics that have driven  the research in the emerging field of antiferromagnetic spintronics  \cite{Jungwirth2016,Baltz2018,Zelezny2018,Gomonay2018,Nemec2018,Smejkal2018}. The development of devices whose operation is based on antiferromagnets was initiated by theoretical predictions  \cite{Shick2010,Zelezny2014} and subsequent experimental demonstrations of electrical detection and manipulation of the antiferromagnetic order by relativistic anisotropic magnetoresistance (AMR) and N\'eel spin-orbit torque (NSOT) effects in metallic antiferromagnets\cite{Park2011b,Marti2014,Wadley2016,Godinho2018}. From the early days of the antiferromagnetic spintronics research, a special attention is paid to complementing these electrical measurements by direct microscopic imaging of the typically multidomain states of the studied antiferromagnets \cite{Wadley2016,Grzybowski2017,Wadley2018,Moriyama2018,Reichlova2019,Gray2019,Bodnar2019,Baldrati2019,Wornle2019,Kaspar2019}. The aim of these microscopies is to elucidate physical mechanisms of the switching which, e.g., in CuMnAs have been associated with the N\'eel vector reorientation induced by the  NSOT, and with electrical or optical pulse-induced quenching into  paidnano-fragmented domain states of the antiferromagnet\cite{Wornle2019,Kaspar2019}. The microscopies are also essential for disentangling potential parasitic non-magnetic contributions to the resistive switching signals, as reported in metal/antiferromagnetic-insulator bilayers\cite{Chiang2019,Zink2019,Zhang2019c,Baldrati2019,Cheng2020,Churikova2020}.  

However, established microscopy techniques for imaging antiferromagnets are rare and rely primarily on large-scale experimental facilities. Among these, X-ray magnetic linear dichroism combined with photoemission electron microscopy  (XMLD-PEEM) \cite{Scholl2000} was used to visualize  the electrical control of the N\'eel vector  in CuMnAs, Mn$_2$Au, or NiO \cite{Wadley2016,Grzybowski2017,Wadley2018,Moriyama2018,Bodnar2019,Baldrati2019}. In CuMnAs, the XMLD-PEEM images of the onset of current-induced NSOT reorientation of the N\'eel vector were directly linked to the onset of the corresponding electrical  readout signals due to AMR \cite{Grzybowski2017,Wadley2018}. 90$^\circ$  N\'eel vector switching was observed by XMLD-PEEM for orthogonal writing currents \cite{Wadley2016,Grzybowski2017} or, via domain wall motion, when reversing the polarity of the writing current \cite{Wadley2018}.
Since XMLD-PEEM is a synchrotron-based technique, more accessible table-top  microscopies are necessary for a systematic exploration of  antiferromagnetic devices. An example here is the NV-diamond magnetometry\cite{Degen2008,Balasubramanian2008} which was recently reported in  antiferromagnetic Cr$_2$O$_3$, BiFeO$_3$, and CuMnAs \cite{Kosub2016,Gross2017,Wornle2019}, and which relies on stray-fields generated by uncompensated magnetic moments. 

In this work we investigate current pulse-induced changes of the domain structure in the compensated collinear antiferromagnet CuMnAs\cite{Wadley2013,Wadley2015a}, focusing on 90$^\circ$  and 180$^\circ$ N\'eel vector switching as well as domain fragmentation. For the microscopic imaging we utilize a thermoelectric response due to the magneto-Seebeck effect (MSE), which is a thermal analog of AMR. The MSE can be applied to the large class of conductive antiferromagnets and is not limited to either uncompensated antiferromagnets that still produce detectable magnetic stray fields, or to systems whose additional broken symmetries allow for the anomalous Nernst effect or the magnetooptical Kerr effect, such as non-collinear antiferromagnets. 

The MSE response is  mapped to a laser-induced localized temperature gradient in the device. A thermoelectric voltage signal is measured across the entire bar device when the scanning probe is placed on top of an antiferromagnetic texture with spatially varying N\'eel vector.

We employ two techniques: The first one is based on the scanning far-field optical microscopy (SFOM) \cite{Weiler2012}, which in combination with  anomalous Nernst or spin-Seebeck thermoelectric response was employed in earlier studies of a non-collinear antiferromagnet Mn$_3$Sn and a metal/antiferromagnetic-insulator bilayer Pt/NiO, respectively  \cite{Reichlova2019,Gray2019}. 
In the second, high-resolution approach we utilize photocurrent nanoscopy in a scattering-type  scanning near-field optical microscope (SNOM) \cite{Pfitzner2018,Woessner2016,Lundeberg2017}.  Here a metal-coated tip of an atomic force microscope (AFM) placed in close proximity to the CuMnAs surface acts as an optical antenna for light focused on the tip. The incident electric field is strongly confined around the tip apex, providing a nanoscale near field point source. Since, to the best of our knowledge, the scanning optical microscopy combined with MSE has not been applied to antiferromagnets prior to our work, we provide  comparisons to images obtained by the established synchrotron XMLD-PEEM technique. 

\vspace*{1cm}

\noindent {\bf Comparison of optical-thermoelectric and  X-ray microscopies of CuMnAs domains}	

\begin{figure}[h!]
\hspace*{-0cm}
\includegraphics[width=.9\textwidth]{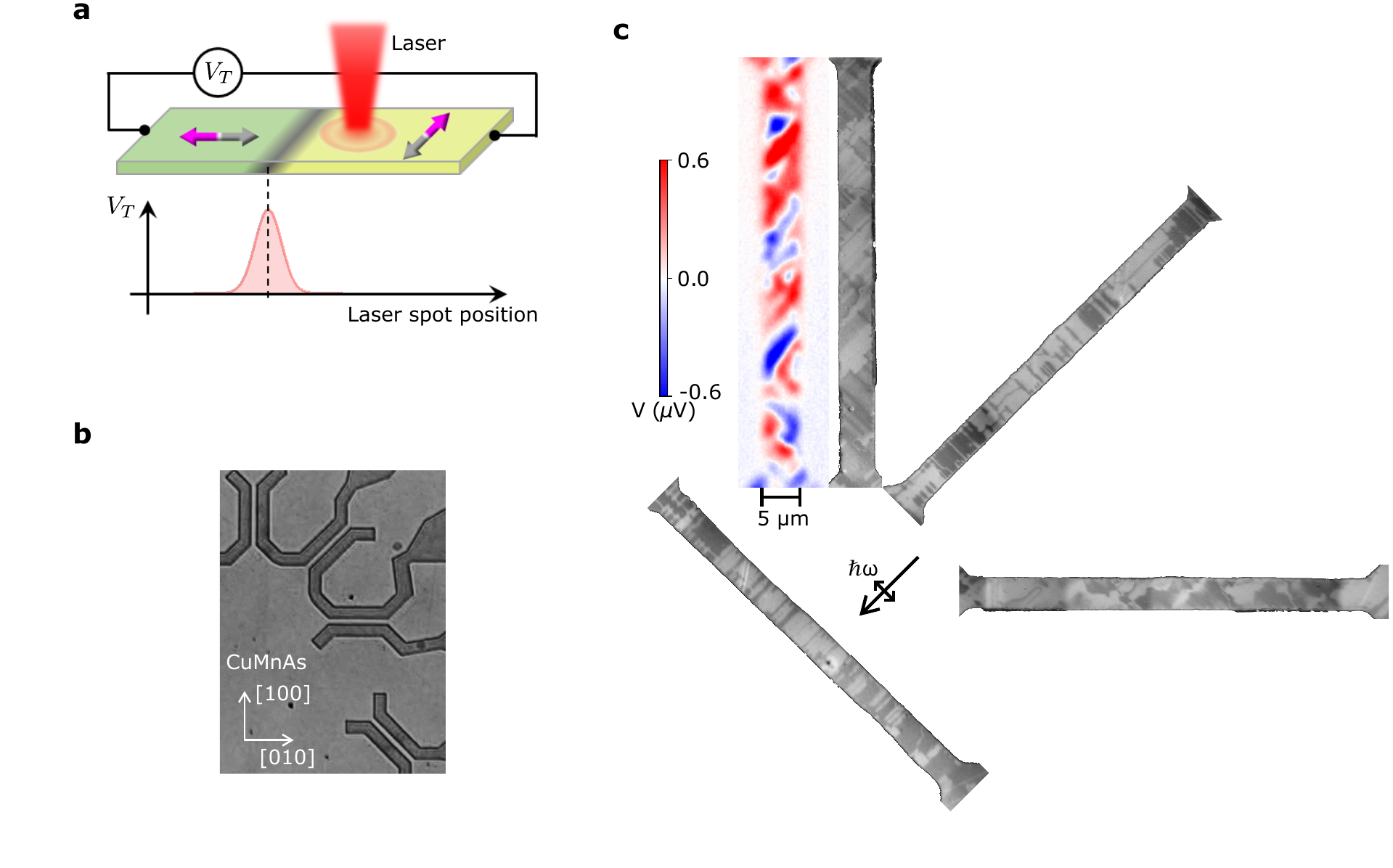}
\caption
{
}
\label{fig-bars}
\end{figure}

\vspace*{.5cm}
In Fig.~\ref{fig-bars}\textbf{a} we illustrate our SFOM-MSE
technique on two neighbouring antiferromagnetic domains separated by a 90$^\circ$ domain wall. We use a 800~nm wavelength  cw-laser beam  of 1 mW power focused to a spot with a full-width at half-maximum (FWHM) of $\approx 1~\mu$m on the surface of the CuMnAs antiferromagnet. The laser spot generates a lateral radially symmetric temperature gradient and we monitor the laser-induced thermoelectric voltage, $V_T$, at the two ends of the bar device. 
Non-zero $V_T$ may occur when the temperature gradient crosses an antiferromagnetic domain boundary, as shown schematically in Fig.~\ref{fig-bars}\textbf{a}.
This is because the N\'eel vector reorients and, therefore, the magneto-Seebeck coefficient changes ~\cite{Krzysteczko2015} so that the net thermoelectric signal does not cancel. As we show in the Supplementary Note~1, we can reproduce the sign and magnitude of the measured  $V_T$ signal with a magneto-Seebeck coefficient  $\Delta S=S_{c}-S_{p}=4 ~\mu$V/K by considering the boundary conditions of our open circuit configuration, thermal conductivities of 200~W/(K$\cdot$m) and  75~W/(K$\cdot$m) for the metallic CuMnAs film and for the insulating GaP substrate, respectively, and by assuming that $50 ~\%$ of the laser  power is  absorbed within the metallic CuMnAs layer. Here $S_{c}$ ($S_{p}$) is the Seebeck coefficient when  the N\'eel vector is collinear (perpendicular) to the temperature gradient.  Note that the calculated maximum temperature rise at 5 ~mW laser power, the highest power used in our SFOM-MSE experiments, is not greater than 6 ~K (see Supplementary Note 1). We also verified that anisotropies of the conductivity, e.g., due to AMR, give a negligible contribution to the  thermoelectric voltage signal.

The optical micrograph in Fig.~\ref{fig-bars}\textbf{b} shows four 50~$\mu$m long and 5~$\mu$m wide bars, which were patterned from a 45~nm thick CuMnAs/GaP epilayer \cite{Wadley2013,Wadley2018} along [100], [1$\bar{1}$0], [110] and [010]  crystallographic axes of CuMnAs. The SFOM-MSE
signals of the four devices  are compared in Fig.~\ref{fig-bars}\textbf{c}  to the XMLD-PEEM measurements taken on the same bars with X-ray polarization $E\parallel [1\bar{1}0]$ crystal axis. The light and dark areas correspond to antiferromagnetic domains with the N\'eel vector oriented perpendicular and parallel to the X-ray polarization, respectively\cite{Wadley2016,Wadley2018}.  
Both domains with orthogonal orientation are found in our 45~nm thick CuMnAs film with the dominant in-plane biaxial magnetic anisotropy \cite{Wadley2018}.

The SFOM-MSE
 and the XMLD-PEEM  measurements show analogous structures of micron-scale domains in each of the four bars. The preferential alignment of the domain walls follows the crystallographic directions of the in-plane square lattice of CuMnAs. This results in the 45$^\circ$ rotation of the preferred domain wall alignment  with respect to the bar edges between the [100] ([010]) and [1$\bar{1}$0] ([110]) bars.

The analogous overall structure of the SFOM-MSE 
and XMLD-PEEM images confirms that the main contribution to the thermoelectric voltage signal comes from the antiferromagnetic texture and the corresponding variation of the magneto-Seebeck coefficient.
Quantitative differences between the two measurements can be ascribed to different lateral resolution and depth sensitivity of the two techniques. The lateral resolution of the XMLD-PEEM in the metallic antiferromagnet CuMnAs is about 50~nm while the resolution of the SFOM-MSE 
is limited by the thermal gradient generated by the $ \sim 1~ \mu$m wide Gaussian shaped laser spot. Regarding the depth sensitivity, the photo-electrons in the XMLD-PEEM are detected only from a few-nm surface layer of the antiferromagnet while the thermoelectric measurements probe the full thickness of the antiferromagnetic film. Note also that the XMLD-PEEM measurements were performed about 10 days before the SFOM-MSE
measurements.

\vspace*{1cm}

\noindent {\bf Optical thermoelectric imaging of the current-induced switching}	

\vspace*{.5cm}

\begin{figure}[h!]
\hspace*{-0cm}\includegraphics[width=.9\textwidth]{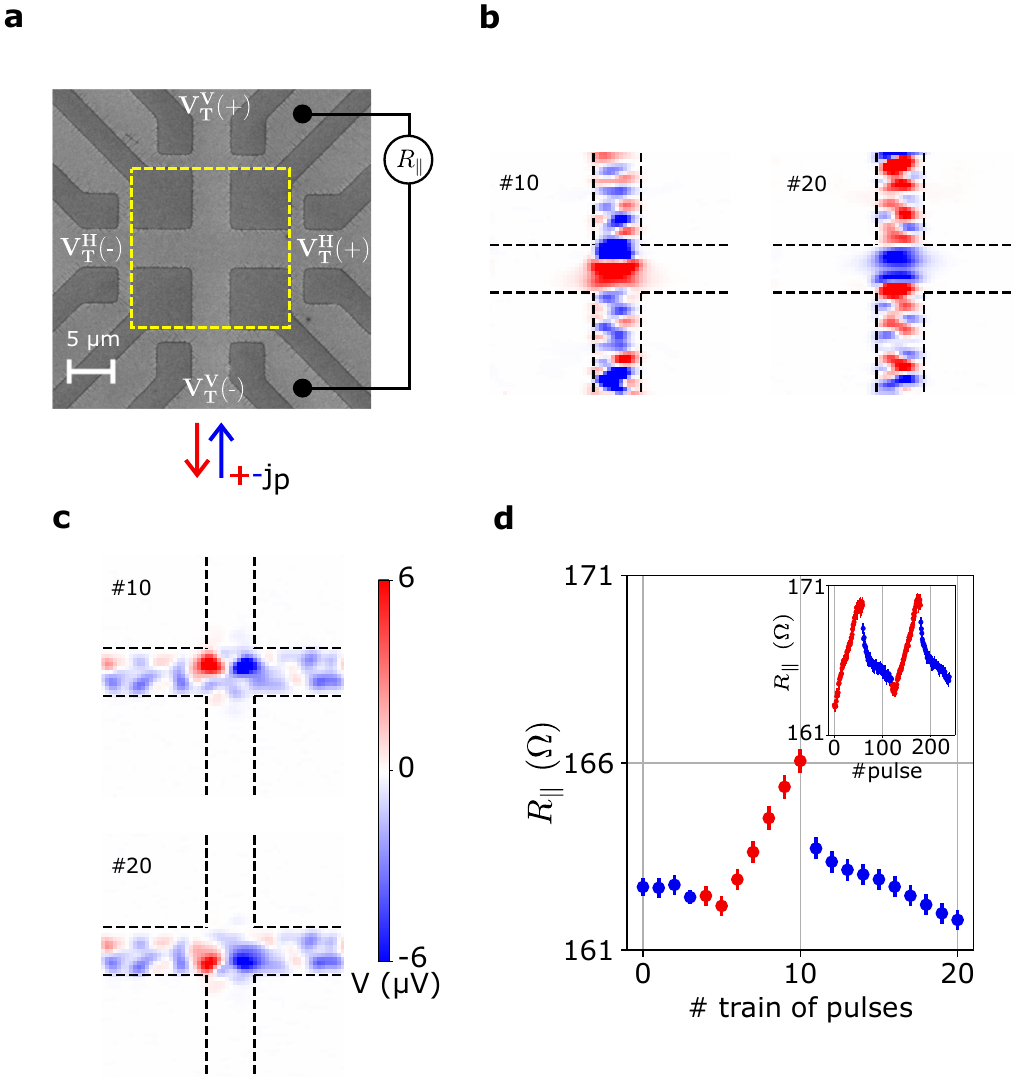}
\caption
{
}
\label{cross_MR}
\end{figure}

We now use the SFOM method  to correlate the local magnetic domain structure to electrical resistance variation after current pulse excitation, which further evidences that the image contrast we detect is indeed of magnetic origin. We simultaneously measure the thermoelectric signals along the vertical and horizontal bars in a symmetric $5 ~\mu$m wide cross bar geometry, shown in Fig.~\ref{cross_MR}\textbf{a} .
The vertical and horizontal SFOM-MSE
voltages $V_{T}^{V} =V_{T}^{V}(+) - V_{T}^{V}(-) $ and $V_{T}^{H}=V_{T}^{H}(+) - V_{T}^{H}(-) $ in Fig.~\ref{cross_MR}\textbf{a}  are recorded while scanning the focused laser spot over the central crossbar structure, highlighted in Fig.~\ref{cross_MR}\textbf{a}  by the dashed yellow rectangle.

Figures~\ref{cross_MR}\textbf{b,c} show the corresponding maps after trains of positive and negative current pulses were applied along the vertical bar with amplitude $|j_p| = 9.6 \times 10^{10}$ A/m$^2$ and duration  $\tau_p = 20~$ms.
Vertical and horizontal thermoelectric signals reflect a complex microscopic domain structure. They appear only when the scanning laser spot illuminates the corresponding 
 bars.  After the applied vertical current pulses, variations of the vertical signal were observed along the entire vertical bar, whereas the horizontal signal changes only in the central overlapping crossbar region. These measurements confirm  Current-pulse-induced switching of the microscopic domain structure since modifications of the thermoelectric signal occur only in areas where the current density of the applied pulse was sufficiently large to trigger the switching \cite{Wadley2018}.

Figure~\ref{cross_MR}\textbf{d}  shows the electrical resistance, $R_{||}$, measured in a 4-point geometry after applying current pulses along the vertical bar. We found that variations in   $R_{||}$ are accompanied with modified SFOM-MSE maps in Figs.~\ref{cross_MR}\textbf{b,c} which are due to the current pulse-induced modification of the domain configuration in the current carrying bar.
$R_{||}$ changes reversibly and reproducibly by applying pulses of opposite polarity, as shown in the inset of Fig.~\ref{cross_MR}\textbf{d}. This is consistent with the NSOT switching mechanism which was identified in the earlier XMLD-PEEM study at comparable amplitudes of the current pulses \cite{Wadley2018}. Note that we observe a change in resistance of about 4\%. This is larger than the expected AMR due to  90$^\circ$ N\'eel vector reorientation inside a domain \cite{Wadley2018,Wang2020} and indicates that additional effects contribute to the variation of $R_{||}$ in our multidomain state.

\begin{figure}[h]
\hspace*{-0cm}\includegraphics[width=.9\textwidth]{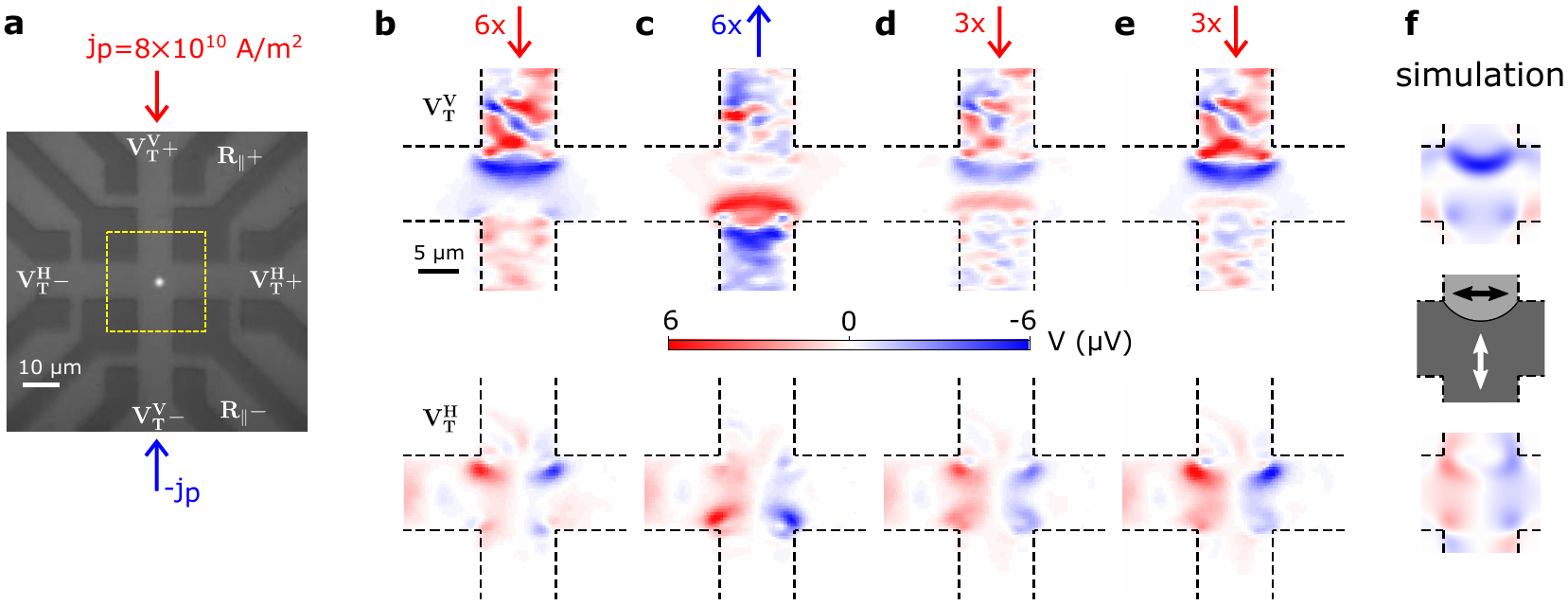}
\caption
{
}
\label{cross_AMT}
\end{figure}

To further evidence the reversible NSOT switching controlled by the current polarity we measure a $10~\mu$m wide symmetric cross bar device, shown in Fig.~\ref{cross_AMT}\textbf{a}. We start by applying 6 positive pulses  along the vertical channel and record the SFOM-MSE map shown in Fig.~\ref{cross_AMT}\textbf{b}. After applying 6 negative pulses we obtain the significantly modified image shown in Fig.~\ref{cross_AMT}\textbf{c}. When applying again 6 positive pulses, we recover the nearly identical original SFOM-MSE map (cf. Figs.~\ref{cross_AMT}\textbf{b,e}).

In Fig.~\ref{cross_AMT}\textbf{f} we simulate the SFOM-MSE measurement considering a realistic domain configuration. We compare the measurements shown in Figs.~\ref{cross_AMT}\textbf{b-e}  with results from self-consistent simulations of the MSE response for the vertical and horizontal bars of a geometrically pinned bubble-shape domain wall in a symmetric cross structure. We consider that due to the higher current density within the bar, domains with their N\'eel vector parallel to the NSOT driving field enlarge their size to gain the effective Zeeman energy\cite{Zelezny2014}. The  domain wall motion remains restricted by geometric pinning at the cross entrance when moving the domain wall towards the cross center \cite{Wunderlich2001,Janda2017}.  The corresponding self-consistently calculated  MSE maps for the vertical bar and the horizontal bar are in good qualitative agreement with our measurements. Details on the simulation  as well as a discussion on the small AMR contributions to the thermoelectric signal can be found in the Supplementary Note~1.

\begin{figure}[h!]
\hspace*{-0cm}\includegraphics[width=.9\textwidth]{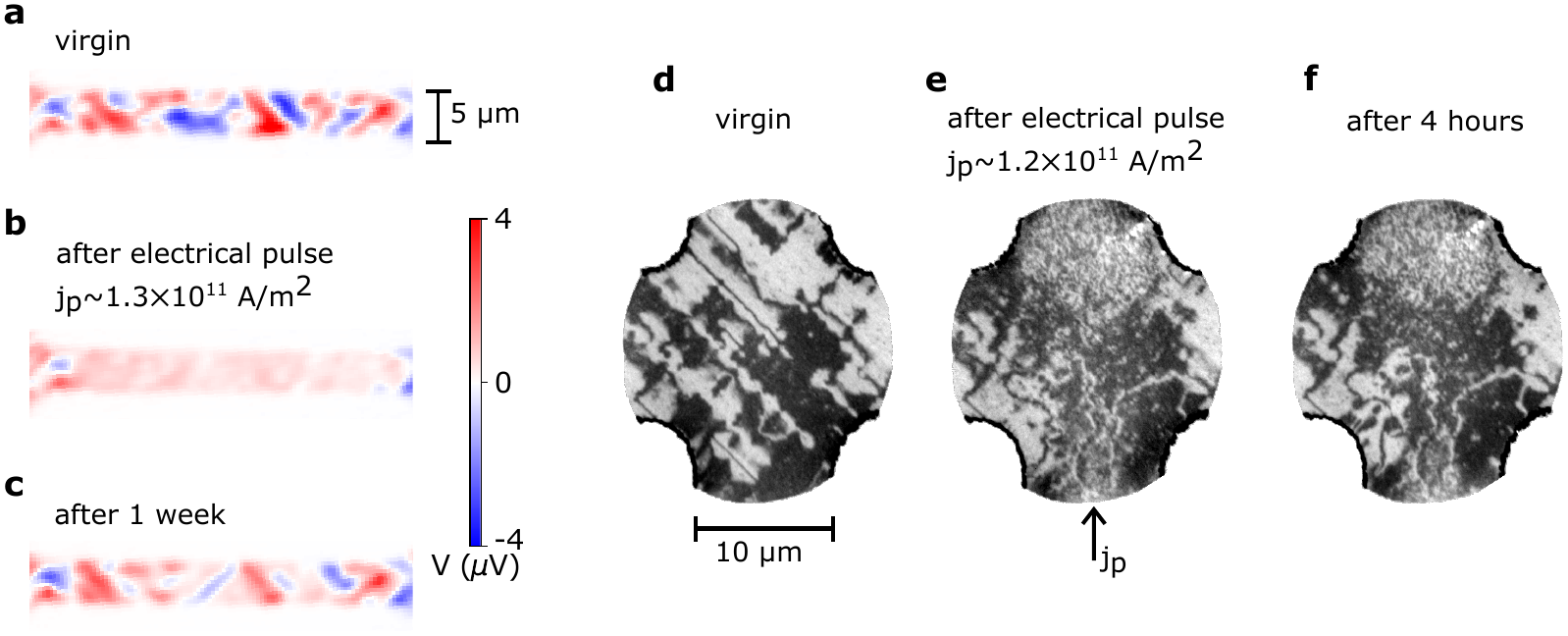}
\caption
{
}
\label{shattering}
\end{figure}

So far we have discussed SFOM-MSE experiments in which electrical pulses of opposite polarity caused reversible N\'eel vector switching via domain wall displacement in the antiferromagnet with micron-scale domains. 
When applying stronger pulses of amplitude $|j_p| = 1.3 \times 10^{11}$ A/m$^2$, we observe diminishing contrast of the  SFOM-MSE signal, as shown in Figs.~\ref{shattering}\textbf{a,b}  for a  $5~\mu$m wide bar.  We ascribe the vanishing SFOM-MSE contrast to a fragmented multi-domain state of the antiferromagnet with sub-micron feature sizes that are significantly smaller  than the extension of the thermal gradient in our SFOM experiment. As a consequence, the net thermoelectric signal from the many domains averages out. 
Acquiring the full SFOM-MSE image after the pulse takes about 30 minutes. For comparison, we show in Figs.~\ref{shattering}\textbf{d-f}  XMLD-PEEM measurements on a similar CuMnAs film and with similar pulse amplitudes, taken a few minutes after the pulse (Fig.~\ref{shattering}\textbf{e})  and again after 4 hours (Fig.~\ref{shattering}\textbf{f}). We see that domains are shattered into a fragmented state with many small sub-micron domains by the  current pulse, consistent with the SFOM image in Fig.~\ref{shattering}\textbf{b}. The large domains on the left and right side of the horizontal channel remained unaffected since they were not exposed to the current pulse. The domain fragmentation in CuMnAs has been  explored in parallel XMLD-PEEM and NV-diamond imaging studies and associated with quenched metastable states which form after pulse-heating the system close to the N\'eel temperature\cite{Wornle2019,Kaspar2019,}. Systematic electrical readout measurements showed that corresponding resistive switching signals can reach giant-magnetoresistance amplitudes of $\sim10-100$\%, i.e., far exceed the signals associated with NSOT-induced N\'eel vector reorientations in the unshattered state with domain sizes in the micron-scale or larger\cite{Wornle2019,Kaspar2019}.

Our laboratory SFOM-MSE technique allows for exploring the relaxation of the metastable fragmented states over long time-scales. Remarkably, when re-measuring the SFOM-MSE signal  one week after the pulse, we find again large-scale SFOM-MSE pattern which resembles the original pattern measured prior to the applied pulse (cf. Figs. 4 \textbf{a, c}). This observation, consistent with the results of the NV-diamond imaging\cite{Wornle2019}, hints to the presence of nucleation and pinning centres  in the CuMnAs film. On the other hand, we also note that the disappearance of the contrast observed in the SFOM-MSE measurements in short times after the pulse 
confirm that potential non-magnetic thermoelectric contributions from defects are small compared to the MSE signal from the antiferromagnetic domains. 

The interpretation of the SFOM-MSE signal in terms of an actual domain structure may only be justified for sizes larger than the spatial resolution, as highlighted in Fig.~\ref{cross_AMT}\textbf{b-e}. However, the feature sizes can be significantly smaller\cite{Wornle2019,Kaspar2019} (see also Figs.~\ref{shattering}\textbf{e,f}). In the following we introduce a high resolution method where we can resolve narrow 180$^\circ$ antiferromagnetic domain walls in a thin CuMnAs film with uniaxial magnetic anisotropy, and observe polarity dependent 180$^\circ$  switching via domain wall displacement.

\vspace*{1cm}

\noindent {\bf High resolution imaging of current-induced displacement of 180$^\circ$  domain walls}

\vspace*{.5cm}

CuMnAs films of thickness $\le 20$~nm exhibit a dominant uniaxial magnetic anisotropy component ascribed to the symmetry breaking between the GaP [110] and [1$\bar{1}$0] axes (CuMnAs [100] and [010] axes)  at  the GaP/CuMnAs interface\cite{Saidl2017,Wadley2015a,Wang2020}. In such films, narrow 180$^\circ$ domain walls separate magnetic domains with reversed N\'eel vectors as shown, e.g., by XMLD-PEEM measurements on a $10~$nm  CuMnAs film in Supplementary Note~2. 

In the following we present MSE measurements on bar devices patterned from a $20~$nm CuMnAs film.  (For further discussion of the uniaxial anisotropy in this film as confirmed by our MSE measurements, see Supplementary Note~2.) 
The detectable MSE-signal in uniaxial films is generated only within the 180$^\circ$ antiferromagnetic domain wall since the two neighbouring domains with opposite N\'eel vectors share the same Seebeck coefficient. In order to image narrow 180$^\circ$ domain walls  we therefore have to generate a thermal gradient with spatial resolution of the order of the domain wall width.  To enhance the spatial resolution we scatter the laser light from a metallic tip, as known from scattering-type SNOM~\cite{Keilmann2004}. This technique allows us to focus light on a spot size of a few tens of nm, only limited by the tip's dimensions \cite{Mastel2018},  and hence to generate a much sharper thermal gradient as compared to the SFOM method.

\begin{figure}[h!] 
\hspace*{-0cm}\includegraphics[width=.9\textwidth]{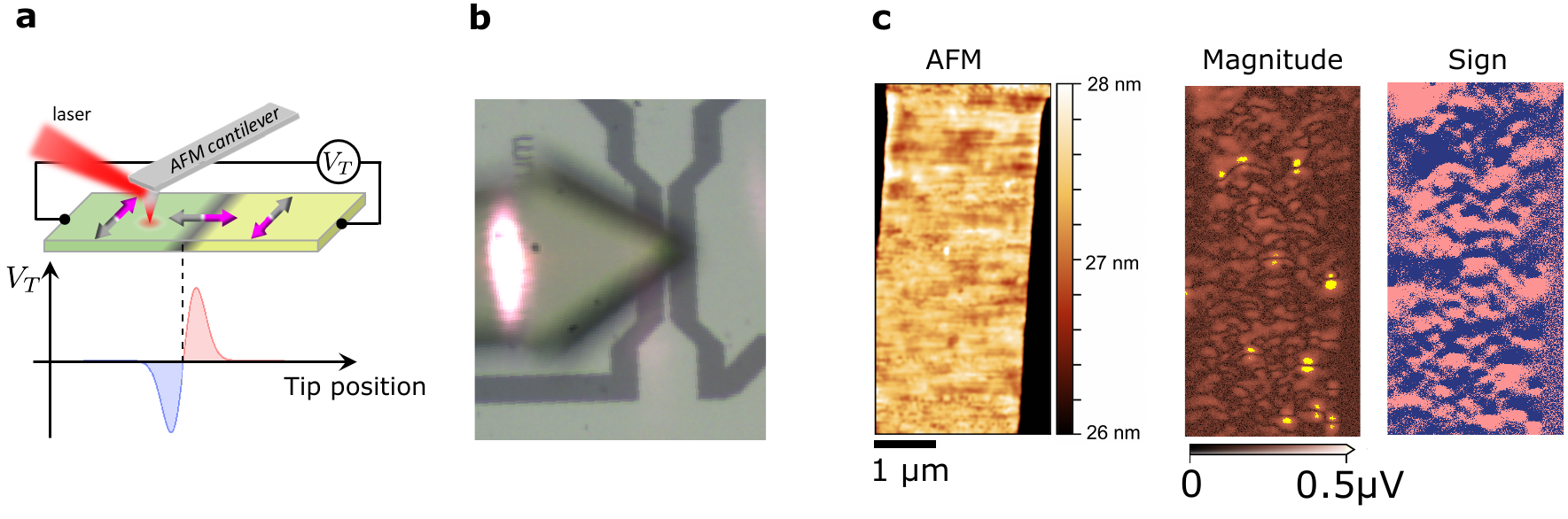}
\caption
{
}
\label{SNOM}
\end{figure}

In Fig.~\ref{SNOM}\textbf{a}  we illustrate our SNOM-MSE technique. The radiation induced temperature profile underneath the tip is indicated by the red spot. The MSE signal appears as two features of the same intensity but opposite polarity when scanning with the AFM tip over the 180$^\circ$ domain wall, since only the variation of the magneto-Seebeck coefficient within the domain wall contributes to the signal. The position where the MSE signal switches sign therefore corresponds to the position of the 180$^\circ$ domain wall.

Figure~\ref{SNOM}\textbf{b}  shows a micrograph of a $2~\mu$m wide CuMnAs bar device below the cantilever with the AFM tip. The thermoelectric voltage, $V_T$, generated in the channel is analyzed by a lock-in amplifier at the AFM tip modulation frequency $\Omega$.
For tip enhanced focusing we use a scattering-type SNOM  operated in the tapping mode. A gold coated Si cantilever  with a typical tip diameter below 50~nm oscillates with an amplitude of 80~nm above the sample surface at its mechanical resonance frequency $\Omega \approx 240\,$kHz. The continuous wave emission of a quantum cascade laser  is focused onto the tip apex which acts as an antenna transmitting a strongly confined near-field to the sample surface.  In contrast to our diffraction-limited SFOM method with  $\lambda = 800~$nm  excitation wavelength, we use here a laser emission with mid-infrared wavelength because the longer wavelength couples more efficiently into the AFM tip and the resolution of this near-field method is not diffraction limited. 

Figure~\ref{SNOM}\textbf{c}  shows, from left to right, the AFM topography image, the magnitude of the thermoelectric voltage $|V_T|$, and its sign $sgn(V_T)$, all detected simultaneously during the SNOM-MSE measurement. As evident from the  comparison between the SNOM-MSE signal and the AFM topography, the majority of the features appearing in the MSE map do not correlate with defects in the topography. We therefore conclude that also in this uniaxial material the contrast originates dominantly from the antiferromagnetic texture. In order to highlight the position of the 180$^\circ$ domain walls, we plot  the absolute value of the measured signal alongside with its polarity. We can then identify the 180$^\circ$ domain walls as meandering zero-signal lines that surround micron-size antiferromagnetic domains.

\begin{figure}[h!] 
\hspace*{-0cm}\includegraphics[width=.9\textwidth]{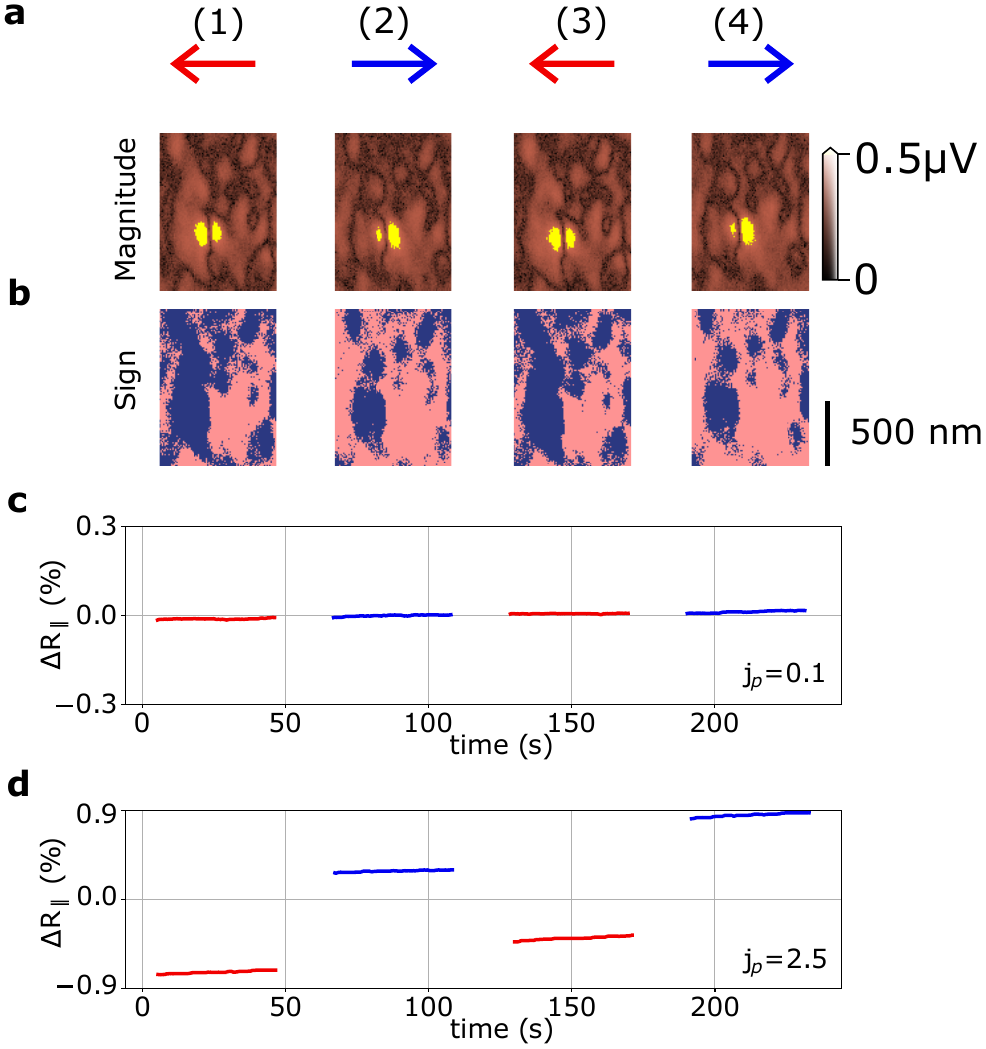}
\caption
{
}

\label{SNOM_pulse}
\end{figure}

In order to investigate the effect of current-induced NSOT on the 180$^\circ$ domain walls we manipulate the magnetic texture by sending current pulses through the bar device, as illustrated in Fig.~\ref{SNOM_pulse}. We apply current pulses of $|j_p| \approx  2.5 \times 10^{11}$ A/m$^2$ with a duration of 1~ms  and with alternating polarity in order to illustrate the reversible switching; the current direction is shown by the red and blue arrows. Note that the onset current amplitude for switching in the 20~nm CuMnAs film is higher than in the above switching experiments in the 45~nm film.  We do not attribute it to the difference of intrinsic properties of the two films. It results from the heat-assisted nature of switching\cite{Olejnik2018,Kaspar2019} and from an interplay of device geometry and heat dissipation during the writing pulse. For ultrashort pulses (with lengths in the ns-scale or smaller), the temperature increase of the CuMnAs device is determined by the energy density delivered by the pulse. Hence,  the onset current density for switching  does not depend on the dimensions of the CuMnAs device\cite{Olejnik2018}. For longer pulses, including those used in the present work, the effect of heat dissipation from the device during the pulse becomes important. Consequently, the current density required to achieve the same switching temperature increases with decreasing film thickness.

In Figs.~\ref{SNOM_pulse}\textbf{a,b} we plot a zoom of the measured $|V_T|$ and $sgn(V_T)$ after applying  a train of 22 current pulses before applying the train of current pulses again with opposite polarity. We found that depending on the polarity of the applied pulses, the antiferromagnetic domains change their size by reversibly displacing domain walls, consistent with the NSOT driven antiferromagnetic domain wall motion \cite{Gomonay2016,Otxoa2020}. The corresponding resistance changes are plotted in Figs.~\ref{SNOM_pulse}\textbf{c,d}.  After applying pulses  of amplitude $|j_p| =  2.5 \times 10^{11}$ A/m$^2$, we observe in Fig.~\ref{SNOM_pulse}\textbf{d} bistable changes of the bar resistance  of the order of $1-2 ~\% $.  In comparison, no changes are observed for  $|j_p| =  0.1 \times 10^{11}$ A/m$^2$, as shown in Fig.~\ref{SNOM_pulse}\textbf{c}. We attribute the resistance variations to magnetic scattering on the domain walls; the AMR contributions from the antiferromagnetic domains can be excluded in the uniaxial film.  More details on the switching as a function of the polarity, number and amplitude of the current pulses by means of principal components analysis can be found in the Supplementary Note~3.

\vspace*{1cm}

\noindent {\bf Conclusions}

We have introduced a  laboratory method for imaging antiferromagnetic domain structure  by mapping the local magneto-Seebeck effect using a far-field or near-field optical scanning approach. In uniaxial CuMnAs, we identify narrow  180$^\circ$ domain walls of sub-micron width and their pulse induced displacements. These reversible,  polarity-dependent modifications of the antiferromagnetic domain maps are consistent with the current-induced NSOT switching mechanism. We link the imaged domain changes to resistive switching signals which we attribute to scattering on the 180$^\circ$ domain walls. In biaxial CuMnAs, we confirm large micron size domains and their Current-pulse-induced modifications. We conclude that AMR from the 90$^\circ$ N\'eel vector reorientation in the antiferromagnetic domains can explain only part of the measured resistance variations. We suggest that magnetic scattering on domain walls gives a strong additional contribution to the observed resistive switching. Apart from the polarity dependent NSOT reorientation of the N\'eel vector at lower pulse amplitudes we also confirm shattering into fragmented metastable multi-domain states with sub-micron feature sizes after applying larger amplitude pulses, and the subsequent relaxation towards the pre-pulsed state of the antiferromagnet.  

\bibliographystyle{naturemag}
\bibliography{Refs,RefsBK}

\begin{thebibliography}{10}
\expandafter\ifx\csname url\endcsname\relax
  \def\url#1{\texttt{#1}}\fi
\expandafter\ifx\csname urlprefix\endcsname\relax\def\urlprefix{URL }\fi
\providecommand{\bibinfo}[2]{#2}
\providecommand{\eprint}[2][]{\url{#2}}

\bibitem{Jungwirth2016}
\bibinfo{author}{Jungwirth, T.}, \bibinfo{author}{Marti, X.},
  \bibinfo{author}{Wadley, P.} \& \bibinfo{author}{Wunderlich, J.}
\newblock \bibinfo{title}{{Antiferromagnetic spintronics}}.
\newblock \emph{\bibinfo{journal}{Nature Nanotechnology}}
  \textbf{\bibinfo{volume}{11}}, \bibinfo{pages}{231--241}
  (\bibinfo{year}{2016}).
\newblock \eprint{1606.04284}.

\bibitem{Baltz2018}
\bibinfo{author}{Baltz, V.} \emph{et~al.}
\newblock \bibinfo{title}{{Antiferromagnetic spintronics}}.
\newblock \emph{\bibinfo{journal}{Reviews of Modern Physics}}
  \textbf{\bibinfo{volume}{90}}, \bibinfo{pages}{015005}
  (\bibinfo{year}{2018}).
\newblock \urlprefix\url{https://doi.org/10.1103/RevModPhys.90.015005}.

\bibitem{Zelezny2018}
\bibinfo{author}{{\v{Z}}elezn{\'{y}}, J.}, \bibinfo{author}{Wadley, P.},
  \bibinfo{author}{Olejn{\'{i}}k, K.}, \bibinfo{author}{Hoffmann, A.} \&
  \bibinfo{author}{Ohno, H.}
\newblock \bibinfo{title}{{Spin transport and spin torque in antiferromagnetic
  devices}}.
\newblock \emph{\bibinfo{journal}{Nature Physics}}
  \textbf{\bibinfo{volume}{14}}, \bibinfo{pages}{220--228}
  (\bibinfo{year}{2018}).

\bibitem{Gomonay2018}
\bibinfo{author}{Gomonay, O.}, \bibinfo{author}{Baltz, V.},
  \bibinfo{author}{Brataas, A.} \& \bibinfo{author}{Tserkovnyak, Y.}
\newblock \bibinfo{title}{{Antiferromagnetic spin textures and dynamics}}.
\newblock \emph{\bibinfo{journal}{Nature Physics}}
  \textbf{\bibinfo{volume}{14}}, \bibinfo{pages}{213--216}
  (\bibinfo{year}{2018}).
\newblock \urlprefix\url{http://dx.doi.org/10.1038/s41567-018-0049-4}.

\bibitem{Nemec2018}
\bibinfo{author}{N{\v{e}}mec, P.}, \bibinfo{author}{Fiebig, M.},
  \bibinfo{author}{Kampfrath, T.} \& \bibinfo{author}{Kimel, A.~V.}
\newblock \bibinfo{title}{{Antiferromagnetic opto-spintronics}}.
\newblock \emph{\bibinfo{journal}{Nature Physics}}
  \textbf{\bibinfo{volume}{14}}, \bibinfo{pages}{229--241}
  (\bibinfo{year}{2018}).
\newblock \urlprefix\url{http://dx.doi.org/10.1038/s41567-018-0051-x}.

\bibitem{Smejkal2018}
\bibinfo{author}{{\v{S}}mejkal, L.} \& \bibinfo{author}{Jungwirth, T.}
\newblock \bibinfo{title}{{Symmetry and topology in antiferromagnetic
  spintronics}}.
\newblock In \bibinfo{editor}{Zang, J.}, \bibinfo{editor}{Cros, V.} \&
  \bibinfo{editor}{Hoffmann, A.} (eds.) \emph{\bibinfo{booktitle}{Topology in
  magnetism}}, \bibinfo{pages}{267--298} (\bibinfo{publisher}{Springer
  International Publishing}, \bibinfo{year}{2018}).
\newblock \urlprefix\url{http://arxiv.org/abs/1804.05628}.

\bibitem{Shick2010}
\bibinfo{author}{Shick, A.~B.}, \bibinfo{author}{Khmelevskyi, S.},
  \bibinfo{author}{Mryasov, O.~N.}, \bibinfo{author}{Wunderlich, J.} \&
  \bibinfo{author}{Jungwirth, T.}
\newblock \bibinfo{title}{{Spin-orbit coupling induced anisotropy effects in
  bimetallic antiferromagnets: A route towards antiferromagnetic spintronics}}.
\newblock \emph{\bibinfo{journal}{Physical Review B}}
  \textbf{\bibinfo{volume}{81}}, \bibinfo{pages}{212409}
  (\bibinfo{year}{2010}).
\newblock \eprint{1002.2151}.

\bibitem{Zelezny2014}
\bibinfo{author}{{\v{Z}}elezn{\'{y}}, J.} \emph{et~al.}
\newblock \bibinfo{title}{{Relativistic n{\'{e}}el-order fields induced by
  electrical current in antiferromagnets}}.
\newblock \emph{\bibinfo{journal}{Physical Review Letters}}
  \textbf{\bibinfo{volume}{113}}, \bibinfo{pages}{157201}
  (\bibinfo{year}{2014}).
\newblock \eprint{1410.8296}.

\bibitem{Park2011b}
\bibinfo{author}{Park, B.~G.} \emph{et~al.}
\newblock \bibinfo{title}{{A spin-valve-like magnetoresistance of an
  antiferromagnet-based tunnel junction}}.
\newblock \emph{\bibinfo{journal}{Nature Materials}}
  \textbf{\bibinfo{volume}{10}}, \bibinfo{pages}{347--351}
  (\bibinfo{year}{2011}).

\bibitem{Marti2014}
\bibinfo{author}{Marti, X.} \emph{et~al.}
\newblock \bibinfo{title}{{Room-temperature antiferromagnetic memory
  resistor}}.
\newblock \emph{\bibinfo{journal}{Nature Materials}}
  \textbf{\bibinfo{volume}{13}}, \bibinfo{pages}{367--374}
  (\bibinfo{year}{2014}).
\newblock \urlprefix\url{http://www.nature.com/articles/nmat3861}.
\newblock \eprint{0402594v3}.

\bibitem{Wadley2016}
\bibinfo{author}{Wadley, P.} \emph{et~al.}
\newblock \bibinfo{title}{{Electrical switching of an antiferromagnet}}.
\newblock \emph{\bibinfo{journal}{Science}} \textbf{\bibinfo{volume}{351}},
  \bibinfo{pages}{587--590} (\bibinfo{year}{2016}).
\newblock \eprint{1503.03765}.

\bibitem{Godinho2018}
\bibinfo{author}{Godinho, J.} \emph{et~al.}
\newblock \bibinfo{title}{{Electrically induced and detected N{\'{e}}el vector
  reversal in a collinear antiferromagnet}}.
\newblock \emph{\bibinfo{journal}{Nature Communications}}
  \textbf{\bibinfo{volume}{9}}, \bibinfo{pages}{4686} (\bibinfo{year}{2018}).
\newblock \urlprefix\url{http://arxiv.org/abs/1806.02795}.
\newblock \eprint{1806.02795}.

\bibitem{Grzybowski2017}
\bibinfo{author}{Grzybowski, M.~J.} \emph{et~al.}
\newblock \bibinfo{title}{{Imaging Current-Induced Switching of
  Antiferromagnetic Domains in CuMnAs}}.
\newblock \emph{\bibinfo{journal}{Physical Review Letters}}
  \textbf{\bibinfo{volume}{118}}, \bibinfo{pages}{057701}
  (\bibinfo{year}{2017}).
\newblock \eprint{1607.08478}.

\bibitem{Wadley2018}
\bibinfo{author}{Wadley, P.} \emph{et~al.}
\newblock \bibinfo{title}{{Current polarity-dependent manipulation of
  antiferromagnetic domains}}.
\newblock \emph{\bibinfo{journal}{Nature Nanotechnology}}
  \textbf{\bibinfo{volume}{13}}, \bibinfo{pages}{362--365}
  (\bibinfo{year}{2018}).
\newblock \urlprefix\url{http://www.nature.com/articles/s41565-018-0079-1}.
\newblock \eprint{arXiv:1711.05146}.

\bibitem{Moriyama2018}
\bibinfo{author}{Moriyama, T.}, \bibinfo{author}{Oda, K.},
  \bibinfo{author}{Ohkochi, T.}, \bibinfo{author}{Kimata, M.} \&
  \bibinfo{author}{Ono, T.}
\newblock \bibinfo{title}{{Spin torque control of antiferromagnetic moments in
  NiO}}.
\newblock \emph{\bibinfo{journal}{Scientific Reports}}
  \textbf{\bibinfo{volume}{8}}, \bibinfo{pages}{14167} (\bibinfo{year}{2018}).
\newblock \urlprefix\url{http://arxiv.org/abs/1708.07682
  http://www.nature.com/articles/s41598-018-32508-w}.
\newblock \eprint{1708.07682}.

\bibitem{Reichlova2019}
\bibinfo{author}{Reichlova, H.} \emph{et~al.}
\newblock \bibinfo{title}{{Imaging and writing magnetic domains in the
  non-collinear antiferromagnet Mn3Sn}}.
\newblock \emph{\bibinfo{journal}{Nature Communications}}
  \textbf{\bibinfo{volume}{10}}, \bibinfo{pages}{5459} (\bibinfo{year}{2019}).
\newblock \urlprefix\url{http://dx.doi.org/10.1038/s41467-019-13391-z
  http://www.nature.com/articles/s41467-019-13391-z}.
\newblock \eprint{1905.13504}.

\bibitem{Gray2019}
\bibinfo{author}{Gray, I.} \emph{et~al.}
\newblock \bibinfo{title}{{Spin Seebeck Imaging of Spin-Torque Switching in
  Antiferromagnetic Pt/NiO Heterostructures}}.
\newblock \emph{\bibinfo{journal}{Physical Review X}}
  \textbf{\bibinfo{volume}{9}}, \bibinfo{pages}{041016} (\bibinfo{year}{2019}).
\newblock \urlprefix\url{https://doi.org/10.1103/PhysRevX.9.041016}.
\newblock \eprint{1810.03997}.

\bibitem{Bodnar2019}
\bibinfo{author}{Bodnar, S.~Y.} \emph{et~al.}
\newblock \bibinfo{title}{Imaging of current induced n\'eel vector switching in
  antiferromagnetic ${\mathrm{mn}}_{2}\mathrm{Au}$}.
\newblock \emph{\bibinfo{journal}{Phys. Rev. B}} \textbf{\bibinfo{volume}{99}},
  \bibinfo{pages}{140409} (\bibinfo{year}{2019}).
\newblock \urlprefix\url{https://link.aps.org/doi/10.1103/PhysRevB.99.140409}.

\bibitem{Baldrati2019}
\bibinfo{author}{Baldrati, L.} \emph{et~al.}
\newblock \bibinfo{title}{{Mechanism of N{\'{e}}el Order Switching in
  Antiferromagnetic Thin Films Revealed by Magnetotransport and Direct
  Imaging}}.
\newblock \emph{\bibinfo{journal}{Physical Review Letters}}
  \textbf{\bibinfo{volume}{123}}, \bibinfo{pages}{177201}
  (\bibinfo{year}{2019}).
\newblock \urlprefix\url{http://arxiv.org/abs/1810.11326
  https://link.aps.org/doi/10.1103/PhysRevLett.123.177201}.
\newblock \eprint{1810.11326}.

\bibitem{Wornle2019}
\bibinfo{author}{W{\"{o}}rnle, M.~S.} \emph{et~al.}
\newblock \bibinfo{title}{{Current-induced fragmentation of antiferromagnetic
  domains}}  (\bibinfo{year}{2019}).
\newblock \urlprefix\url{http://arxiv.org/abs/1912.05287}.
\newblock \eprint{1912.05287}.

\bibitem{Kaspar2019}
\bibinfo{author}{Ka{\v{s}}par, Z.} \emph{et~al.}
\newblock \bibinfo{title}{{High resistive unipolar-electrical and fs-optical
  switching in a single-layer antiferromagnetic memory}} \bibinfo{pages}{1--22}
  (\bibinfo{year}{2019}).
\newblock \urlprefix\url{http://arxiv.org/abs/1909.09071}.
\newblock \eprint{1909.09071}.

\bibitem{Chiang2019}
\bibinfo{author}{Chiang, C.~C.}, \bibinfo{author}{Huang, S.~Y.},
  \bibinfo{author}{Qu, D.}, \bibinfo{author}{Wu, P.~H.} \&
  \bibinfo{author}{Chien, C.~L.}
\newblock \bibinfo{title}{{Absence of Evidence of Electrical Switching of the
  Antiferromagnetic N{\'{e}}el Vector}}.
\newblock \emph{\bibinfo{journal}{Physical Review Letters}}
  \textbf{\bibinfo{volume}{123}}, \bibinfo{pages}{227203}
  (\bibinfo{year}{2019}).
\newblock
  \urlprefix\url{https://link.aps.org/doi/10.1103/PhysRevLett.123.227203}.

\bibitem{Zink2019}
\bibinfo{author}{Zink, B.}
\newblock \bibinfo{title}{{The Heat in Antiferromagnetic Switching}}.
\newblock \emph{\bibinfo{journal}{Physics}} \textbf{\bibinfo{volume}{12}},
  \bibinfo{pages}{134} (\bibinfo{year}{2019}).

\bibitem{Zhang2019c}
\bibinfo{author}{Zhang, P.}, \bibinfo{author}{Finley, J.},
  \bibinfo{author}{Safi, T.} \& \bibinfo{author}{Liu, L.}
\newblock \bibinfo{title}{{Quantitative Study on Current-Induced Effect in an
  Antiferromagnet Insulator/Pt Bilayer Film}}.
\newblock \emph{\bibinfo{journal}{Physical Review Letters}}
  \textbf{\bibinfo{volume}{123}}, \bibinfo{pages}{247206}
  (\bibinfo{year}{2019}).
\newblock \urlprefix\url{http://arxiv.org/abs/1907.00314}.
\newblock \eprint{1907.00314}.

\bibitem{Cheng2020}
\bibinfo{author}{Cheng, Y.}, \bibinfo{author}{Yu, S.}, \bibinfo{author}{Zhu,
  M.}, \bibinfo{author}{Hwang, J.} \& \bibinfo{author}{Yang, F.}
\newblock \bibinfo{title}{{Electrical Switching of Tristate Antiferromagnetic
  N{\'{e}}el Order in alpha Fe2O3 Epitaxial Films}}.
\newblock \emph{\bibinfo{journal}{Physical Review Letters}}
  \textbf{\bibinfo{volume}{124}}, \bibinfo{pages}{027202}
  (\bibinfo{year}{2020}).
\newblock
  \urlprefix\url{http://arxiv.org/abs/1906.04694{\%}0Ahttps://link.aps.org/doi/10.1103/PhysRevLett.124.027202}.
\newblock \eprint{1906.04694}.

\bibitem{Churikova2020}
\bibinfo{author}{Churikova, A.} \emph{et~al.}
\newblock \bibinfo{title}{{Non-magnetic origin of spin Hall
  magnetoresistance-like signals in Pt films and epitaxial NiO/Pt bilayers}}.
\newblock \emph{\bibinfo{journal}{Applied Physics Letters}}
  \textbf{\bibinfo{volume}{116}}, \bibinfo{pages}{022410}
  (\bibinfo{year}{2020}).
\newblock \urlprefix\url{http://aip.scitation.org/doi/10.1063/1.5134814}.

\bibitem{Scholl2000}
\bibinfo{author}{Scholl, A.}
\newblock \bibinfo{title}{{Observation of Antiferromagnetic Domains in
  Epitaxial Thin Films}}.
\newblock \emph{\bibinfo{journal}{Science}} \textbf{\bibinfo{volume}{287}},
  \bibinfo{pages}{1014--1016} (\bibinfo{year}{2000}).
\newblock
  \urlprefix\url{https://www.sciencemag.org/lookup/doi/10.1126/science.287.5455.1014}.

\bibitem{Degen2008}
\bibinfo{author}{Degen, C.~L.}
\newblock \bibinfo{title}{{Scanning magnetic field microscope with a diamond
  single-spin sensor}}.
\newblock \emph{\bibinfo{journal}{Applied Physics Letters}}
  \textbf{\bibinfo{volume}{92}}, \bibinfo{pages}{243111}
  (\bibinfo{year}{2008}).

\bibitem{Balasubramanian2008}
\bibinfo{author}{Balasubramanian, G.} \emph{et~al.}
\newblock \bibinfo{title}{{Nanoscale imaging magnetometry with diamond spins
  under ambient conditions}}.
\newblock \emph{\bibinfo{journal}{Nature}} \textbf{\bibinfo{volume}{455}},
  \bibinfo{pages}{648--651} (\bibinfo{year}{2008}).
\newblock \urlprefix\url{http://www.nature.com/articles/nature07278}.

\bibitem{Kosub2016}
\bibinfo{author}{Kosub, T.} \emph{et~al.}
\newblock \bibinfo{title}{{Purely antiferromagnetic magnetoelectric random
  access memory}}.
\newblock \emph{\bibinfo{journal}{Nature Communications}}
  \textbf{\bibinfo{volume}{8}}, \bibinfo{pages}{13985} (\bibinfo{year}{2017}).
\newblock \urlprefix\url{http://www.nature.com/articles/ncomms13985}.
\newblock \eprint{1611.07027}.

\bibitem{Gross2017}
\bibinfo{author}{Gross, I.} \emph{et~al.}
\newblock \bibinfo{title}{{Skyrmion morphology in ultrathin magnetic films}}.
\newblock \emph{\bibinfo{journal}{Physical Review Materials}}
  \textbf{\bibinfo{volume}{2}}, \bibinfo{pages}{024406} (\bibinfo{year}{2018}).
\newblock
  \urlprefix\url{http://arxiv.org/abs/1709.06027{\%}0Ahttp://dx.doi.org/10.1103/PhysRevMaterials.2.024406
  https://link.aps.org/doi/10.1103/PhysRevMaterials.2.024406}.
\newblock \eprint{1709.06027}.

\bibitem{Wadley2013}
\bibinfo{author}{Wadley, P.} \emph{et~al.}
\newblock \bibinfo{title}{{Tetragonal phase of epitaxial room-temperature
  antiferromagnet CuMnAs}}.
\newblock \emph{\bibinfo{journal}{Nature Communications}}
  \textbf{\bibinfo{volume}{4}}, \bibinfo{pages}{2322} (\bibinfo{year}{2013}).
\newblock \eprint{1402.3624}.

\bibitem{Wadley2015a}
\bibinfo{author}{Wadley, P.} \emph{et~al.}
\newblock \bibinfo{title}{{Antiferromagnetic structure in tetragonal CuMnAs
  thin films}}.
\newblock \emph{\bibinfo{journal}{Scientific Reports}}
  \textbf{\bibinfo{volume}{5}}, \bibinfo{pages}{17079} (\bibinfo{year}{2015}).

\bibitem{Weiler2012}
\bibinfo{author}{Weiler, M.} \emph{et~al.}
\newblock \bibinfo{title}{{Local charge and spin currents in magnetothermal
  landscapes}}.
\newblock \emph{\bibinfo{journal}{Physical Review Letters}}
  \textbf{\bibinfo{volume}{108}}, \bibinfo{pages}{106602}
  (\bibinfo{year}{2012}).
\newblock \eprint{arXiv:1110.3981}.

\bibitem{Pfitzner2018}
\bibinfo{author}{Pfitzner, E.} \emph{et~al.}
\newblock \bibinfo{title}{{Near-field magneto-caloritronic nanoscopy on
  ferromagnetic nanostructures}}.
\newblock \emph{\bibinfo{journal}{AIP Advances}} \textbf{\bibinfo{volume}{8}},
  \bibinfo{pages}{125329} (\bibinfo{year}{2018}).

\bibitem{Woessner2016}
\bibinfo{author}{Woessner, A.} \emph{et~al.}
\newblock \bibinfo{title}{{Near-field photocurrent nanoscopy on bare and
  encapsulated graphene}}.
\newblock \emph{\bibinfo{journal}{Nature Communications}}
  \textbf{\bibinfo{volume}{7}}, \bibinfo{pages}{10783} (\bibinfo{year}{2016}).
\newblock \eprint{1508.07864}.

\bibitem{Lundeberg2017}
\bibinfo{author}{Lundeberg, M.~B.} \emph{et~al.}
\newblock \bibinfo{title}{{Thermoelectric detection and imaging of propagating
  graphene plasmons}}.
\newblock \emph{\bibinfo{journal}{Nature Materials}}
  \textbf{\bibinfo{volume}{16}}, \bibinfo{pages}{204--207}
  (\bibinfo{year}{2017}).

\bibitem{Krzysteczko2015}
\bibinfo{author}{Krzysteczko, P.}, \bibinfo{author}{Hu, X.},
  \bibinfo{author}{Liebing, N.}, \bibinfo{author}{Sievers, S.} \&
  \bibinfo{author}{Schumacher, H.~W.}
\newblock \bibinfo{title}{{Domain wall magneto-Seebeck effect}}.
\newblock \emph{\bibinfo{journal}{Physical Review B}}
  \textbf{\bibinfo{volume}{92}}, \bibinfo{pages}{1--5} (\bibinfo{year}{2015}).
\newblock \eprint{1412.8289}.

\bibitem{Wang2020}
\bibinfo{author}{Wang, M.} \emph{et~al.}
\newblock \bibinfo{title}{{Spin flop and crystalline anisotropic
  magnetoresistance in CuMnAs}}.
\newblock \emph{\bibinfo{journal}{Physical Review B}}
  \textbf{\bibinfo{volume}{101}}, \bibinfo{pages}{094429}
  (\bibinfo{year}{2020}).
\newblock \urlprefix\url{http://arxiv.org/abs/1911.12381
  https://link.aps.org/doi/10.1103/PhysRevB.101.094429}.
\newblock \eprint{1911.12381}.

\bibitem{Wunderlich2001}
\bibinfo{author}{Wunderlich, J.} \emph{et~al.}
\newblock \bibinfo{title}{{Influence of geometry on domain wall propagation in
  a mesoscopic wire}}.
\newblock \emph{\bibinfo{journal}{IEEE Trans. Mag.}}
  \textbf{\bibinfo{volume}{37}}, \bibinfo{pages}{2104--2107}
  (\bibinfo{year}{2001}).

\bibitem{Janda2017}
\bibinfo{author}{Janda, T.} \emph{et~al.}
\newblock \bibinfo{title}{{Inertial displacement of a domain wall excited by
  ultra-short circularly polarized laser pulses}}.
\newblock \emph{\bibinfo{journal}{Nature Communications}}
  \textbf{\bibinfo{volume}{8}}, \bibinfo{pages}{15226} (\bibinfo{year}{2017}).
\newblock \urlprefix\url{http://www.nature.com/articles/ncomms15226}.

\bibitem{}
\bibinfo{title}{{John Robert Schrieffer - Theory of Superconductivity (1999,
  Westview Press)}}.

\bibitem{Saidl2017}
\bibinfo{author}{Saidl, V.} \emph{et~al.}
\newblock \bibinfo{title}{{Optical determination of the N{\'{e}}el vector in a
  CuMnAs thin-film antiferromagnet}}.
\newblock \emph{\bibinfo{journal}{Nature Photonics}}
  \textbf{\bibinfo{volume}{11}}, \bibinfo{pages}{91--96}
  (\bibinfo{year}{2017}).
\newblock \eprint{1608.01941}.

\bibitem{Keilmann2004}
\bibinfo{author}{Keilmann, F.} \& \bibinfo{author}{Hillenbrand, R.}
\newblock \bibinfo{title}{{Near-field microscopy by elastic light scattering
  from a tip}}.
\newblock \emph{\bibinfo{journal}{Philosophical Transactions of the Royal
  Society of London. Series A: Mathematical, Physical and Engineering
  Sciences}} \textbf{\bibinfo{volume}{362}}, \bibinfo{pages}{787--805}
  (\bibinfo{year}{2004}).
\newblock
  \urlprefix\url{https://royalsocietypublishing.org/doi/10.1098/rsta.2003.1347}.

\bibitem{Mastel2018}
\bibinfo{author}{Mastel, S.} \emph{et~al.}
\newblock \bibinfo{title}{{Understanding the Image Contrast of Material
  Boundaries in IR Nanoscopy Reaching 5 nm Spatial Resolution}}.
\newblock \emph{\bibinfo{journal}{ACS Photonics}} \textbf{\bibinfo{volume}{5}},
  \bibinfo{pages}{3372--3378} (\bibinfo{year}{2018}).

\bibitem{Olejnik2018}
\bibinfo{author}{Olejn{\'{i}}k, K.} \emph{et~al.}
\newblock \bibinfo{title}{{Terahertz electrical writing speed in an
  antiferromagnetic memory}}.
\newblock \emph{\bibinfo{journal}{Science Advances}}
  \textbf{\bibinfo{volume}{4}}, \bibinfo{pages}{eaar3566}
  (\bibinfo{year}{2018}).
\newblock
  \urlprefix\url{http://advances.sciencemag.org/lookup/doi/10.1126/sciadv.aar3566}.
\newblock \eprint{1711.08444}.

\bibitem{Gomonay2016}
\bibinfo{author}{Gomonay, O.}, \bibinfo{author}{Jungwirth, T.} \&
  \bibinfo{author}{Sinova, J.}
\newblock \bibinfo{title}{{High Antiferromagnetic Domain Wall Velocity Induced
  by N{\'{e}}el Spin-Orbit Torques}}.
\newblock \emph{\bibinfo{journal}{Physical Review Letters}}
  \textbf{\bibinfo{volume}{117}}, \bibinfo{pages}{017202}
  (\bibinfo{year}{2016}).
\newblock \eprint{1602.06766}.

\bibitem{Otxoa2020}
\bibinfo{author}{Otxoa, R.~M.}, \bibinfo{author}{Roy, P.~E.},
  \bibinfo{author}{Rama-Eiroa, R.}, \bibinfo{author}{Giuslienko, K.~Y.} \&
  \bibinfo{author}{Wunderlich, J.}
\newblock \bibinfo{title}{{Walker-like domain wall breakdown in layered
  antiferromagnets driven by staggered spin-orbit fields}}
  (\bibinfo{year}{2020}).
\newblock \urlprefix\url{http://arxiv.org/abs/2002.03332}.
\newblock \eprint{2002.03332}.

\end{thebibliography}


\begin{thebibliography}{1}
\expandafter\ifx\csname url\endcsname\relax
  \def\url#1{\texttt{#1}}\fi
\expandafter\ifx\csname urlprefix\endcsname\relax\def\urlprefix{URL }\fi
\providecommand{\bibinfo}[2]{#2}
\providecommand{\eprint}[2][]{\url{#2}}

\bibitem{Wadley2013}
\bibinfo{author}{Wadley, P.} \emph{et~al.}
\newblock \bibinfo{title}{{Tetragonal phase of epitaxial room-temperature
  antiferromagnet CuMnAs}}.
\newblock \emph{\bibinfo{journal}{Nature Communications}}
  \textbf{\bibinfo{volume}{4}}, \bibinfo{pages}{2322} (\bibinfo{year}{2013}).
\newblock \eprint{1402.3624}.

\bibitem{Wunderlich2001}
\bibinfo{author}{Wunderlich, J.} \emph{et~al.}
\newblock \bibinfo{title}{{Influence of geometry on domain wall propagation in
  a mesoscopic wire}}.
\newblock \emph{\bibinfo{journal}{IEEE Trans. Mag.}}
  \textbf{\bibinfo{volume}{37}}, \bibinfo{pages}{2104--2107}
  (\bibinfo{year}{2001}).

\bibitem{Janda2017}
\bibinfo{author}{Janda, T.} \emph{et~al.}
\newblock \bibinfo{title}{{Inertial displacement of a domain wall excited by
  ultra-short circularly polarized laser pulses}}.
\newblock \emph{\bibinfo{journal}{Nature Communications}}
  \textbf{\bibinfo{volume}{8}}, \bibinfo{pages}{15226} (\bibinfo{year}{2017}).
\newblock \urlprefix\url{http://www.nature.com/articles/ncomms15226}.

\bibitem{Wang2020}
\bibinfo{author}{Wang, M.} \emph{et~al.}
\newblock \bibinfo{title}{{Spin flop and crystalline anisotropic
  magnetoresistance in CuMnAs}}.
\newblock \emph{\bibinfo{journal}{Physical Review B}}
  \textbf{\bibinfo{volume}{101}}, \bibinfo{pages}{094429}
  (\bibinfo{year}{2020}).
\newblock \urlprefix\url{http://arxiv.org/abs/1911.12381
  https://link.aps.org/doi/10.1103/PhysRevB.101.094429}.
\newblock \eprint{1911.12381}.

\bibitem{Saidl2017}
\bibinfo{author}{Saidl, V.} \emph{et~al.}
\newblock \bibinfo{title}{{Optical determination of the N{\'{e}}el vector in a
  CuMnAs thin-film antiferromagnet}}.
\newblock \emph{\bibinfo{journal}{Nature Photonics}}
  \textbf{\bibinfo{volume}{11}}, \bibinfo{pages}{91--96}
  (\bibinfo{year}{2017}).
\newblock \eprint{1608.01941}.

\bibitem{Wadley2015a}
\bibinfo{author}{Wadley, P.} \emph{et~al.}
\newblock \bibinfo{title}{{Antiferromagnetic structure in tetragonal CuMnAs
  thin films}}.
\newblock \emph{\bibinfo{journal}{Scientific Reports}}
  \textbf{\bibinfo{volume}{5}}, \bibinfo{pages}{17079} (\bibinfo{year}{2015}).

\bibitem{Pearson1901}
\bibinfo{author}{Pearson, K.}
\newblock \bibinfo{title}{{LIII. On lines and planes of closest fit to systems
  of points in space}}.
\newblock \emph{\bibinfo{journal}{The London, Edinburgh, and Dublin
  Philosophical Magazine and Journal of Science}} \textbf{\bibinfo{volume}{2}},
  \bibinfo{pages}{559--572} (\bibinfo{year}{1901}).
\newblock \urlprefix\url{https://doi.org/10.1080/14786440109462720}.

\bibitem{Jolliffe2020}
\bibinfo{author}{Jolliffe, I.~T.}
\newblock \emph{\bibinfo{title}{{Principal Components Analysis}}}
  (\bibinfo{publisher}{Springer}, \bibinfo{year}{2020}).

\end{thebibliography}

\newpage

\noindent {\bf Acknowledgments}
B.K., A.H and G.U. acknowledge funding from the EMPIR programme (JRP ADVENT) co-financed by the Participating States and from the European UnionÕs Horizon 2020 research and innovation programme. 
E.P. and J.H. acknowledge funding from the DFG-project HE 2063/5-1.
T.Ja., J.G.,ZT.S., H.R.,V.N., K.O., J.W. and T.Ju.  acknowledge funding from Ministry of Education of the Czech Republic Grant No. LM2018110 and LNSM-LNSpin, of the Czech Science Foundation Grant No. 19-28375X, and  for the EU FET Open RIA Grant No. 766566. 
J.G., V.N., K.O., and J.W.  acknowledges funding from the ERC Synergy grant no. 610115. We acknowledge Diamond Light Source for time on Beamline I06 under proposal nos. SI16376 and SI20793.
\vspace*{1cm}

\noindent {\bf Author contributions}

T.Ja., P.N., and J.W. planned and prepared the SFOM-MSE measurements, which T.Ja. and J.G. carried out. 
G.U., E.P. , A.H., and B.K. planned and prepared the SNOM-MSE measurements, which J.W., G.U., E.P., A.H., J.G., H.R., T.M., and B.K. carried out.
J.G., H.R., J.W., and K.O. planned and  performed the electrical pulsing experiments.
S.R., O.J.A, J.S.C., P.W., K.W.E., S.S.D., F.M. planned and performed the XMLD-PEEM measurements.
T.O. performed the simulations. 
V.N., R.P.C., and P.W. grew and characterized the CuMnAs films.
Z.S. designed and fabricated the devices. 
J.W., B.K., J.G., T.Ja., T.O., T.M., E.P., and  T.Ju. wrote the paper . 
All authors discussed the results and contributed to the manuscript.
B.K. and  J.W. conceived and planned the study. 

\vspace*{1cm}

\noindent {\bf Additional information}
Corresponding authors correspondence to T. Janda, J. Godinho, B. Kaestner, and J. Wunderlich.
\vspace*{1cm}

\noindent {\bf Methods}

{\bfseries Sample fabrication}
For patterning our samples we used  standard electron beam lithography on an PMMA resist film after cleaning the surface of our CuMnAs wafers with acetone. 
After removing the Al/AlO$_{x}$ capping layer using diluted TMAH developer the individual  devices were defined by etching insulating trenches using a mixture of H$_{2}$SO$_{4}$, C$_{4}$H$_{6}$O$_{6}$, H$_{2}$O$_{2}$, and DI H$_{2}$O. 
Bonding contacts were made using a lift-off process following the Cr(3$\,$nm)Au(80$\,$nm) evaporation.

{\bfseries SFOM-MSE technique} The laser beam emited by a Ti:Sapphire continuous-wave (cw) laser (Spectra Physics, model 3900S) tuned to a wavelength of $800\,$nm is focused into a spot-size of $\approx1.5\,\mu$m full width at half maximum (FWHM) by an objective lens (Mitutoyo Plan Apo 20). The data in Fig. 1 was measured with a laser power of $1\,$mW and the data in Figs. 2--4 with a power of $5\,$mW. Since the MSE signal is linear in the laser power (as confirmed by a test measurement, not shown here) the only effect of a larger laser power is a correspondingly larger MSE voltage and a higher signal to noise ratio. Scanning of the laser spot across the sample surface is achieved by moving the objective lens with a 3D piezo-positioner (Newport, NPXYZ100SG-D). The laser beam is modulated at a frequency of $ \Omega  \approx 1.7\,$kHz by an optical chopper and from the measured and amplified MSE voltage the signal component at the chopper-frequency is extracted by a lock-in amplifier.

{\bfseries SNOM-MSE technique} The emission of a quantum cascade laser (QCL, 28$\,$mW at $\lambda \approx 10\,\mu$m, MIRcat, Daylight Solutions Inc., CA, USA) was focussed onto the metallic tip apex of the metal-coated Si cantilever (neaspec nano-FTIR Scanning Probes). Focussing and scanning of the tip was performed using a commercial scattering-type scanning nearfield optical microscopy instrument (Neasnom, by Neaspec GmbH). The tip-mediated electric response of the sample was amplified by low-noise voltage preamplifier (Stanford Research SR 560, gain $ =5 \times 10^3$), and further demodulated at the tip-modulation frequency $\Omega$ and its higher harmonics with the lock-in amplifier of the Neasnom instrument. Both, amplitude and phase, were recorded while scanning the sample surface.

\newpage

\noindent {\bf Figure Captions}
\vspace*{1cm}

\noindent {\bf Figure~\ref{fig-bars}} 
{\bf Comparison between laboratory optical and synchrotron x-ray images of the domain structure in bar-shaped patterned antiferromagnetic CuMnAs.} {\bf a,} Schematics of the measurement setup for the laboratory SFOM-MSE technique (top panel). A focused laser beam creates a local thermal gradient. When scanned over an antiferromagnetic (purple and grey arrows) texture, the in-plane components of the thermal gradient generate a voltage across the bar due to the MSE (bottom panel). {\bf b,}
Optical micrograph of four 50~$\mu$m long and 5~$\mu$m wide bars patterned from a 45 nm thick CuMnAs epilayer along [100], [1$\bar{1}$0], [110], and [010] crystallographic axes of CuMnAs. {\bf c,} Comparison between SFOM-MSE and XMLD-PEEM  measurements in the four microbars. Antiferromagnetic domain structure is observed by XMLD-PEEM for X-ray polarization $E\parallel [1\bar{1}0]$ crystal axis. The single- and double-headed arrows in {\bf c} indicate the in-plane projection of the X-ray propagation vector and the X-ray polarization vector, respectively. The light (dark) contrast corresponds to antiferromagnetic domains with the N\'eel vector oriented perpendicular (parallel) to the X-ray polarization.

\vspace*{1cm}

\noindent {\bf Figure~\ref{cross_MR}} 
{\bf Current-pulse-induced modification of the domain structure  detected by MSE scans and compared to AMR measurements.} {\bf a,} SEM micrograph of a 5$\mu$m wide cross bar patterned from a 45 nm thick CuMnAs epilayer. The MSE scans have been performed within the the area of $25 \times 25 ~\mu$m$^2$, indicated by the yellow dashed line. {\bf b,} MSE signal measured along the vertical bar after 7 trains of positive pulses (left) followed by 10 trains of negative pulses (right). Each train of pulses contains 6 individual pulses {\bf c,} MSE signal simultaneously measured along the horizontal bar. {\bf d,}  Corresponding variation of the magnetoresistance measured in a 4-point geometry along the vertical bar. Red (blue) data points correspond to resistance measurements after current pulses of positive (negative) polarity with $|j_p| = 9.6 \times 10^{10}$ A/m$^2$ and $\tau_p = 2~$ms.

\vspace*{1cm}

\noindent {\bf Figure~\ref{cross_AMT}} 
{\bf Reproducibility of antiferromagnetic texture after pulsing with alternating polarity}
{\bf a,} Optical micrograph of a $10~\mu$m wide cross bar, showing the measurement contacts geometry used. {\bf b - e} Sequence of MSE maps of the vertical thermoelectric voltage and horizontal thermoelectric voltage measured simultaneously for alternating pulses. The MSE scans were performed within the area highlighted by the yellow dashed line in {\bf a}.
{\bf f,} Simulated MSE maps of $V^{V}_{T}$  (upper graph) and $V^{H}_{T}$  (lower graph) for a domain configuration of a geometrically pinned domain wall (middle schematics) by taking into account the experimental conditions of a focused laser spot with Gaussian profile of $1.5~\mu$m FWHM,  $5~$mW laser power and by assuming a magneto-Seebeck coefficient of  $\Delta S=4 ~\mu$V/K.

\vspace*{1cm}

\noindent {\bf Figure~\ref{shattering}} 
{\bf Shattering of magnetic domains after an electrical pulse.} {\bf a,} SFOM-MSE scan of a $5~\mu$m wide bar device prior to any electrical pulse. {\bf b,} After a pulse with a current density of $1.3\times10^{11}$~A/m$^2$ the contrast is lost with the main features absent. The loss of contrast is ascribed to the shattering of the magnetic domains into smaller domains, which become significantly smaller than the spatial resolution of the SFOM-MSE. 
Hence, the total SFOM-MSE signal over many domain walls averages nearly to zero.
{\bf c,} The SFOM-MSE signal one week after the electrical pulse, where a similar pattern of large domains has reappeared and is resembling the initial state in {\bf a}. {\bf d,} XMLD-PEEM measurements prior the pulse, {\bf e}, just after the pulse  with a current density of $1.2 \times10^{11}$~A/m$^2$ {\bf e}, and {\bf f,} 4 hours after the pulse.

\vspace*{1cm}

\noindent {\bf Figure~\ref{SNOM}} 
{\bf SNOM-MSE scan of a bar device patterned from a uniaxial CuMnAs layer.} {\bf a,} Schematics of the SNOM-MSE setup.  The thermal gradient is created when a metal-coated AFM tip interacts with an infra-red laser, inducing an optical near-field at the apex of the tip. A $180^\circ$ domain wall appears in the MSE maps as two features of opposite sign together, as illustrated in the bottom panel. {\bf b,} Micrograph of the scanned bar-device, where the AFM tip is also visible. {\bf c,} From left to right: topography map, magnitude of the MSE signal and its sign.

\vspace*{1cm}

\noindent {\bf Figure~\ref{SNOM_pulse}} 
{\bf Reversible switching measured in SNOM-MSE.} {\bf a,} Maps of the magnitude of the MSE signal, $|V_T|$, after applying current pulses of amplitude $|j_p| =  2.5 \times 10^{11}$ A/m$^2$ in opposite directions indicated by the red and blue arrow on top. {\bf b,} Maps of the corresponding $sgn(V_T)$ values. {\bf c,d,}  Resistance variations associated with the corresponding switching states for low and high current densities. Current density values shown are in $10^{11}$~A/m$^2$.

\newpage

\newpage

\newpage

\newpage

\newpage

\newpage




\end{document}


\title{Supplementary information: Magneto-Seebeck 
microscopy of domain switching in collinear antiferromagnet CuMnAs}
\author{T.~Janda}
\affiliation{Faculty of Mathematics and Physics, Charles University, Ke Karlovu 3, 121 16 Prague 2, Czech Republic}
\affiliation{Institute of Experimental and Applied Physics, University of  Regensburg, Universit\"atsstra{\ss}e 31, 93051 Regensburg}
\author{J.~Godinho}
\affiliation{Institute of Physics, Czech Academy of Sciences, Cukrovarnick\'a 10, 162 00, Praha 6, Czech Republic}
\affiliation{Faculty of Mathematics and Physics, Charles University, Ke Karlovu 3, 121 16 Prague 2, Czech Republic}
\author{T.~Ostatnicky}
\affiliation{Faculty of Mathematics and Physics, Charles University, Ke Karlovu 3, 121 16 Prague 2, Czech Republic}
\author{E.~Pfitzner}
\affiliation{Department of Physics, Freie Universit\"at Berlin, 14195 Berlin, Germany}
\author{G.~Ulrich}
\affiliation{Physikalisch-Technische Bundesanstalt, 38116 Braunschweig and 10587 Berlin, Germany}
\author{A.~Hoehl}
\affiliation{Physikalisch-Technische Bundesanstalt, 38116 Braunschweig and 10587 Berlin, Germany}
\author{S.~Reimers}
\affiliation{School of Physics and Astronomy, University of Nottingham, Nottingham NG7 2RD, United Kingdom}
\author{Z.~\v{S}ob\'a\v{n}} 
\affiliation{Institute of Physics, Czech Academy of Sciences, Cukrovarnick\'a 10, 162 00, Praha 6, Czech Republic}
\author{T.~Metzger} 
\affiliation{Institute of Experimental and Applied Physics, University of  Regensburg, Universit\"atsstra{\ss}e 31, 93051 Regensburg}
\author{H.~Reichlov\'a}
\affiliation{Institut fuer Festkoerper- und Materialphysik, Technische Universit\"at Dresden, 01062 Dresden, Germany}
\author{V.~Nov\'ak} 
\affiliation{Institute of Physics, Czech Academy of Sciences, Cukrovarnick\'a 10, 162 00, Praha 6, Czech Republic}
\author{R.~P.~Campion}
\affiliation{School of Physics and Astronomy, University of Nottingham, Nottingham NG7 2RD, United Kingdom}
\author{J.~Heberle}
\affiliation{Department of Physics, Freie Universit\"at Berlin, 14195 Berlin, Germany}
\author{P.~Wadley}
\affiliation{School of Physics and Astronomy, University of Nottingham, Nottingham NG7 2RD, United Kingdom}
\author{K.~W.~Edmonds}
\affiliation{School of Physics and Astronomy, University of Nottingham, Nottingham NG7 2RD, United Kingdom}
\author{S.~S.~Dhesi}
\affiliation{Diamond Light Source, Chilton, Didcot, United Kingdom}
\author{F.~Maccherozzi}
\affiliation{Diamond Light Source, Chilton, Didcot, United Kingdom}
\author{R.~M.~Otxoa}
\affiliation{Hitachi Cambridge Laboratory, Cambridge CB3 0HE, United Kingdom}
\affiliation{Donostia International Physics Center, 20018 San Sebasti\'an, Spain}
\author{P.~E.~Roy}
\affiliation{Hitachi Cambridge Laboratory, Cambridge CB3 0HE, United Kingdom}
\author{K.~Olejn\'ik} 
\affiliation{Institute of Physics, Czech Academy of Sciences, Cukrovarnick\'a 10, 162 00, Praha 6, Czech Republic} 
\author{P.~N\v{e}mec}
\affiliation{Faculty of Mathematics and Physics, Charles University, Ke Karlovu 3, 121 16 Prague 2, Czech Republic}
\author{T.~Jungwirth}
\affiliation{Institute of Physics, Czech Academy of Sciences, Cukrovarnick\'a 10, 162 00, Praha 6, Czech Republic} 
\affiliation{School of Physics and Astronomy, University of Nottingham, Nottingham NG7 2RD, United Kingdom}
\author{B.~Kaestner}
\affiliation{Physikalisch-Technische Bundesanstalt, 38116 Braunschweig and 10587 Berlin, Germany}
\author{J.~Wunderlich}
\affiliation{Institute of Experimental and Applied Physics, University of  Regensburg, Universit\"atsstra{\ss}e 31, 93051 Regensburg}
\affiliation{Institute of Physics, Czech Academy of Sciences, Cukrovarnick\'a 10, 162 00, Praha 6, Czech Republic}

\maketitle

\noindent {\bf Supplementary Note 1: Simulation of the magneto-Seebeck signal in a cross-bar structure}	

\vspace*{.5cm}

In Fig. 1 {\bf a} of the main text we sketch two antiferromagnetic domains in a biaxial system with different Seebeck coefficient and separated by a $90^\circ$ domain wall. If a laser spot, which generates a lateral radially symmetric temperature gradient, crosses the domain wall, a single uni-polar voltage peak with  the polarity depending on whether the domain with larger or with smaller Seebeck coefficient is placed between domain wall and, e.g., the positive electrical contact. 
In Fig. 3 of the main text we compare the measured thermoelectric voltage signal with a theoretical simulation. The simulation is performed by first deriving the temperature profile in our patterned CuMnAs films  \cite{Wadley2013}, which consist of a 45\,nm thick structured   CuMnAs film, a  0.5\,mm GaP substrate  and a Cu block, on which the GaP substrate is thermally anchored and which acts as a constant temperature reservoir with a temperature of $T=  300$\,K.
In a second simulation step, the thermoelectric magneto-Seebeck signal is calculated for a plausible domain configuration consisting of a single arc-shaped  domain wall which extends into the central cross area due to the current-pulse-induced domain wall motion. The arc-shaped domain wall is a consequence of the  geometric pinning at the corners of the crossbar \cite{Wunderlich2001,Janda2017}.

In our experiments, the surface of the CuMnAs layer is locally heated by a cw-laser beam (1.5\,$\mu$m FWHM) of 5\,mW laser power. We assume that the total absorbed power in the CuMnAs layer is only 50\,\% of the nominal laser power, i.e. 2.5\,mW, because part of the radiation is reflected from the CuMnAs surface. Since CuMnAs is metallic, we assume a thermal conductivity of 200\,W/Km which is a typical value for metals. The numerically derived temperature profile is shown in the Supplementary Fig.~\ref{profile}{\bf a}. It is important to note that the  excess temperature profile due to the laser heating remains approximately Gaussian with only slightly larger FWHM of 2.3~$\mu$m (instead of 1.5~$\mu$m of the focused laser spot) and with a temperature enhancement of  6\,K peak value. For confirmation,  we have calculated the temperature profile from an analytical solution of the heat transport equation by considering an infinitely thin CuMnAs layer. With the analytical approach, we  obtained very similar values  of 7.1\,K peak excess temperature and 2.3~$\mu$m FWHM. 

Since the electromotive force generated by the Seebeck effect depends on the temperature gradient, we compare both the laser-spot-induced temperature gradient in the direction perpendicular to the surface, Fig.~\ref{profile}{\bf b},  and in the sample plane, Fig.~\ref{profile}{\bf c}. Interestingly, these gradients are of similar size but the voltage signal that can be measured between the external contacts is only generated by the in-plane temperature gradient, since the N\'eel vectors are oriented everywhere within the sample plane.

The electromotive force generated by the Seebeck effect  $\vec{E}_{\rm emf}={-}S\nabla T $, where  $S$ is the Seebeck tensor, is not necessarily aligned with the temperature gradient $\nabla T $ since anisotropies of the magneto-Seebeck effect can be generated by the symmetry-breaking magnetic order. For the sake of simplicity, we chose the main axis of the Seebeck tensor $S$ to be aligned collinear with the N\'eel vector.  
The electric current density $ \vec{j}$ results as a combined action of thermally generated electromotive force and the electrostatic Coulomb force:
\begin{equation} 
  \vec{j}={-}\sigma\big[\nabla\varphi+S\nabla T\big] \,,\label{eq_j1}
\end{equation}
where $\varphi$ is the electric potential and $\sigma$ is the electric conductivity tensor. 
Both the matrix elements of the electric conductivity and the Seebeck tensor are connected with the magnetic order due to AMR and MSE, respectively, and both can contribute to the thermoelectric voltage signal when local variations of $\sigma$ and $S$ are located within the temperature gradient. 

For the open circuit geometry of our experiments, the stationary solution is derived from the continuity equation, with  the boundary conditions
\begin{eqnarray}
  {-}\nabla\cdot\vec{j}&=&0 \,,\label{eq_divj}\\
  \vec{j}\cdot\vec{n}&=&0 \hbox{ on the surface}\,,\label{eq_bc}
\end{eqnarray}
where $\vec{n}$ is the unit vector locally normal to the sample surface. Since the electrostatic potential $\varphi$ is determined by the charge density $\rho$ in the sample volume, we solve Eq. (\ref{eq_divj}) with the boundary condition (\ref{eq_bc}) by finding the appropriate charge density $\rho$ and considering Eq. (\ref{eq_j1}) and the Poisson's equation $\epsilon\bigtriangleup\varphi=\rho$. The symbol $\epsilon$ denotes here the sample's electric permittivity and $\bigtriangleup$ is the Laplace operator. The resulting two-terminal voltage is then the difference of the electrostatic potentials at the external contact positions.
By comparing the magnitude of the measured thermoelectric signal with our numerical simulation we estimate a  magneto-Seebeck effect in our material of $\Delta S=S_c - S_p=4~\mu$V/K. $S_c$ and  $S_p$ are the Seebeck coefficients if the  N\'eel vector is collinear or perpendicular to the temperature gradient.

We now discuss the contribution of conductivity variations due to the AMR effect on the thermoelectric voltage signal $V_T$. In Fig.~\ref{anisotropy}{\bf a}, we show the relative change of $V_T$ as a function of AMR $=( \sigma_c - \sigma_p) / ( \sigma_c + \sigma_p)$ with $\sigma_{\rm \parallel}$ ($\sigma_p$) being the component of the conductivity in the direction collinear (perpendicular) to the N\'eel vector.  We relate thermoelectric voltage signals affected by conductivity variations due to the AMR   \cite{Wang2020}  of up to $\pm 10\%$  with the thermoelectric signal generated without AMR.
 The simulation is performed for a 10\,$\mu$m wide bar device with the laser spot of 1.5\,$\mu$m FWHM focused in the center of the bar and exactly on top of a  $90^{\circ}$ domain wall  of 100\,nm width. 
 Since in our more complex cross-bar structures domain walls can be aligned under arbitrary angles with respect to the bar orientation, we also present simulations for domain walls which cross the bar under different crossing angles $\psi$.  From our simulation we can conclude that even unrealistically  large conductivity variations of  $\pm 10\%$ AMR affect only insignificantly the MSE response.

In Fig.~\ref{anisotropy}{\bf b} we show the thermoelectric signal generated in the environment of a $180^\circ$ domain wall if the spatial extension of a non-zero temperature gradient, as shown in Fig.~\ref{profile}{\bf c}, is of the order of the domain wall width, as it is the case in our high resolution SNOM-MSE technique of Figs.\,\,5 and 6 of the main text. In the simulation we compare again the thermoelectric voltage signals affected by conductivity variations due to AMR contributions of $\pm 10\%$  with the MSE signal generated at isotropic conductivity (no AMR). Also here, the thermoelectric signal is  only insignificantly affected by the AMR.

Finally, we present  simulation of the thermoelectric signal in a 2~$\mu$m wide bar device where the N\'eel vector is oriented $45^\circ$ from the bar axis and the thermoelectric signal is generated by the transverse MSE at the bar boundaries. Here, the radial symmetry of the temperature gradient is broken and, therefore, the transverse magneto-Seebeck effect generates a thermoelectric voltage signal with opposite sign on both sides of the bar. In Fig.~\ref{anisotropy}{\bf c}, we plot the simulated transverse MSE signal in the case of isotropic conductivity and $1.5\,\mu$m wide laser spot, i.e., the signal which is expected in the low-resolution SFOM-MSE measurement. Together with the zero-AMR curve we also show the relative change of the MSE signal when AMR of $\pm 10\%$ is taken into account. Fig.~\ref{anisotropy}{\bf d} shows analogous calculations performed for a sharper illumination profile ($100\,$nm FWHM), which is the case relevant for the high-resolution SNOM-MSE measurements. Note that in Figs.~\ref{anisotropy}{\bf c} and~\ref{anisotropy}{\bf d} the relative change of the MSE signal by AMR is enhanced by a factor of 3000 and 500, respectively, i.e., we find again no significant effect of AMR.

Our simulations presented in Fig.~\ref{anisotropy} show that conductivity variations due to the AMR of $\pm 10\%$ do only insignificantly modify the thermoelectric voltage signal $V_T$. We therefore neglect magnetoresistive contributions in the following calculations. 

In the case of homogeneous and isotropic Seebeck coefficient (without the magnetic order) and far from sample boundaries, the equation (\ref{eq_divj}) could be immediately solved by setting $\varphi={-}ST$ which then leads to $\vec{j}=0$. The presence of the bar boundaries would be then compensated by stationary surface charges which in turn do not influence the aforementioned solution and thus the measured voltage. Nonzero anisotropy of the Seebeck coefficient, on the other hand, is unambiguously connected with the macroscopic currents caused by the temperature gradient. We show this effect in our crossbar device geometry in Fig.~\ref{potential} where we plot the calculated electrostatic potential for two distinct positions of the laser spot and we also plot the accompanying current density. We have chosen the laser spot position in the center of a domain (Fig.~\ref{potential}{\bf a},{\bf b}), and on the top of a $90^\circ$ domain wall (Fig.~\ref{potential}{\bf c},{\bf d}). An interesting property of the Seebeck coefficient anisotropy is, that the electrostatic potential is strongly affected by inhomogeneities in the material: either the presence of its surfaces, domain walls, imperfections etc. The reason stems from the spatial restriction of the current density which results in the accumulation of the electric charge which then becomes a source of the Coulomb force. This effect is detectable in our experiments as a voltage measured between contacts not only in the vicinity of the domain walls but also close to sample edges and corners in the cross-shaped samples.

\vspace*{1cm}

\noindent {\bf Supplementary Note 2: Transverse Seebeck effect in CuMnAs with uniaxial magnetic anisotropy}

\vspace*{.5cm}

So far we have assumed that the thermoelectric voltage arises from a radially symmetric temperature gradient in the sample plane in an area with a non-uniform Seebeck coefficient, e.g., at and around domain walls. If the Seebeck coefficient is uniform, all thermal electricity contributions average to zero.
However, if the in-plane component of the temperature gradient is not radially symmetric, thermal current contributions do not average to zero also in regions with uniform Seebeck coefficient. Such a situation can be realized when the laser spot is focused on one of the two bar edges of the device. 
In this case, the  in-plane temperature gradient is no longer radially symmetric, since the heat flow outside the bar differs from the flow inside the bar.  As a result, net temperature gradients point from the boundaries to the center of the bar.
If the Seebeck effect is isotropic, the potential difference appears, however, only perpendicular to the bar, so that the potentials at the external bar contacts remain unaffected.
On the other hand, if the Seebeck effect is anisotropic, as in case of the magneto-Seebeck effect, a transverse voltage signal arises if the magnetic order breaks the transverse symmetry of the bar. This is the case if the  N\'eel vector is neither oriented collinear nor  perpendicular to the bar orientation. In the following we make use of this consideration to give evidence for the dominant uniaxial magnetic anisotropy of our $20\,$nm thin film with $180^\circ$ domain walls. 

CuMnAs films of thickness $\le 20\,$nm exhibit a dominant uniaxial magnetic anisotropy component ascribed to the symmetry breaking between the GaP [110] and [1$\bar{1}$0] axes (CuMnAs [100] and [010] axes)  at  the GaP/CuMnAs interface\cite{Saidl2017,Wadley2015a,Wang2020}. In such films, narrow 180$^\circ$ domain walls separate magnetic domains with reversed N\'eel vectors as shown, e.g., by XMLD-PEEM measurements on a $10\,$nm thin CuMnAs film in Fig.~\ref{uniaxial}.

As illustrated in Fig.~\ref{transverse}{\bf a}, we maximize the transverse magneto-Seebeck effect by fixing the N\'eel vector at $45^\circ$ with respect to the bar orientation. Since the MSE signal does not change under N\'eel vector reversal the transverse MSE can give an indication about the uniaxial magnetic anisotropy of the antiferromagnetic film.
The transversal thermoelectric MSE signal at the edge is then constant along the entire bar extension and of reversed sign at the opposite bar edge. Variations to the constant signal would appear only if the temperature gradient coincides with magnetic domain walls and if the temperature gradient extension is as small or smaller than the domain wall width.

In contrary,  at biaxial multi-domain configurations, edge MSE signals  appear with opposite polarities along one side of the bar corresponding to alternating domains with  $45^\circ$  and  $-45^\circ$  N\'eel vector, or, for small domain size, the thermoelectric signal averages to zero.
 In Figs.~\ref{anisotropy}{\bf c} and~\ref{anisotropy}{\bf d}, we have simulated for a wide and a narrow temperature profile, respectively, the thermoelectric MSE response if a laser spot is swept from one side to the other side of a single domain bar with the N\'eel vector oriented at $45^\circ$ with respect to the bar edges.
 
In Fig.~\ref{transverse}{\bf b}, we show thermoelectric SNOM- and SFOM-MSE measurements on a bar patterned along $45^\circ$ with respect to the expected uniaxial magnetic anisotropy axis of a $20\,$nm thin CuMnAs.  Remarkably, this $45^\circ$ oriented bar generates  transverse MSE signals with the expected reversed signal polarities at opposite bar edges and extending along the entire bar. On the other hand, no transverse MSE signal has been detected on bars patterned from the same thin CuMnAs film but oriented along $0^\circ$  and $90^\circ$. SNOM-MSE measurements on these $0^\circ$ and $90^\circ$ bars are presented in Figs. 5 and 6 of the main text.  Comparing SFOM- and SNOM-MSE measurements, as shown in  Fig.~\ref{transverse}{\bf b}, reveals a much higher spatial resolution of the SNOM-MSE maps. In case of the SFOM measurements,  where  the signal is simultaneously generated and averaged within the spatially extended temperature gradient of the micron-size laser spot, we observe smooth  transverse MSE signals. In contrary, the fine structure of the more complex SNOM-MSE signals is attributed to the local thermoelectric signals originating from individual narrow domain walls. 

\vspace*{1cm}

\noindent {\bf Supplementary Note 3: Current-pulse-induced domain wall displacement: principal components analysis}	

\vspace*{.5cm}

Antiferromagnetic domains can be altered reversibly by current-pulse-induced displacement of domain walls (DWs) as shown in Fig. 6 of the main text in case of CuMnAs with uniaxial magnetic anisotropy. 
We identify the corresponding  $180^\circ$ DWs by mapping the thermoelectric voltage signal, $V_T$, at the terminals of the CuMnAs bar while scanning an illuminated AFM tip across the bar structure. The sharp lines of sign changes indicate the position of the DWs. In Figs. 6{\bf a} and 6{\bf b} of the main text we therefore plot magnitude and sign of $V_T$ separately.
Changing the sizes of domains corresponds then to expanding and shrinking of the respective blue and red regions in the sign$(V_T)$ - map. The DW displacement depends on the current density, polarity and also on the number of pulses. However, it is not straightforward to analyze the complex DW displacements of the multiple domains within the bar. In order to quantify amount and reversibility we apply the method of principal components analysis (PCA)~\cite{Pearson1901, Jolliffe2020}.

The entire bar and the current pulse polarities are shown schematically in the supplementary Fig.~\ref{PCA}{\bf a}. 
We record  the $V_T$-maps only from a small area of $3 \times 1~\mu$m$^2$ as shown in Fig.~\ref{PCA}{\bf b}. 
In order to ensure that we analyse data taken from exactly the same area, we compare and match AFM phase maps which were recorded together with the corresponding SNOM-MSE maps.
The PCA of the $V_T$-map was performed on the grey shaded region of $\Delta x  \times \Delta y = 700 \times 100\,$nm$^2$ shown in Fig.~\ref{PCA}{\bf c} which contains a single DW, visible as a sharp horizontal line of sign change. In this area, we expect domain wall displacements only along vertical directions. We therefore average the data along horizontal lines and further discretize along the $x$-axis into  $p = 100$ segments, "samples", to obtain the resulting functions $\overline{V}_T(n, x_i)$ with $i = 1, \dots p$, as plotted in Fig.~\ref{Specs}{\bf a}. The Index $n$ stands for the state after the $n^\mathrm{th}$ pulse sequence. The PCA was performed by first standardizing the data $\overline{V}_T(n, x_i)$ with the 9-dimensional column vectors ${\bf v}_{(x_i)} = \left[\overline{V}_T(1, x_i), \overline{V}_T(2, x_i), \dots, \overline{V}_T(9, x_i)\right]^T_{(x_i)}$ into the data matrix ${\bf S}$ with each row vector  ${\bf s}_{(n)} = (s_{x_1},s_{x_2}, \dots, s_{x_p})_{(n)} =\left[ \frac{{\bf v}_{(x_i)} - \mu\left({\bf v}_{(x_i)}\right)}{\sigma\left({\bf v}_{(x_i)}\right)}\right]^T$, with $ \mu\left({\bf v}_{(x_i)}\right)$ and $\sigma\left({\bf v}_{(x_i)}\right)$ the mean and standard deviation over the 9 elements of the column vector ${\bf v}_{(x_i)}$, respectively. The PCA transformation maps each row vector ${\bf s}_{(n)}$ to a new vector of so called principal component scores ${\bf t}_{(n)}$ by 
\begin{equation}
 t_{k (n)} = {\bf s}_{(n)} \cdot {\bf w}_{(k)} \quad \mathrm{with} \quad n = 1, \dots , 9 \quad k = 1, \dots, l \ll p
\end{equation}
where ${\bf w}_{(k)} = ( w_{x_1 (k)},  w_{x_2 (k)}, \dots ,  w_{x_p (k)})$ are the eigenvectors of the correlation matrix 
\begin{equation}
	\frac{\mathrm{Cov}\left( {\bf v}_{x_i} , {\bf v}_{x_j} \right)}{  
	\sigma\left({\bf v}_{(x_i)}\right) \sigma\left({\bf v}_{(x_j)}\right)} 
	\quad \mathrm{with} \quad i,j = 1, \dots, p
\end{equation}
with Cov() being the covariance between the column vectors ${\bf v}_{(x_i)}$. The scores are sorted such that the strongest variation with $n$ is associated with the smallest $k$, so that the number $l$ of required eigenvectors to describe the variation of the measurements is ideally much less than $p$. The variance ratios of the scores in the above measurement are 0.68, 0.16, and 0.06 for k=1, 2 and 3, respectively, showing that the main variation is contained in the first 2 principal components. 

More specifically, $\overline{V}_T(n, x_i)$ can now be decomposed into a set of basis functions, $v^{k}_T(x_i)$, as
\begin{equation}
	\overline{V}_T(n, x_i) = \frac{1}{9}\sum_{n = 1}^{9}  \overline{V}_T(n, x_i) - \sum_{k = 1}^{l} t_{k(n)} \, v^{k}_T(x_i)
\end{equation}
with $v^{k}_T(x_i) =  w_{x_i (k)} \, \sigma\left({\bf v}_{(x_i)}\right)$, which are plotted in Fig.~\ref{Specs}{\bf b} for $k = 1 \dots 3$. 

The scores $t_{k(n)}$ of the first principal component $k = 1$ containing the highest spatial variance has been used to evaluate the change of the magnetic texture with pulsing. It is shown as bar graph in Fig.~\ref{PCA}{\bf c}, aligned with the current direction shown by the red and blue arrows. With this analysis the alternating behaviour for pulses in opposite directions can be seen clearly. Moreover, the values obtained for 22 consecutive pulses are higher than the values for 6 pulses, showing the proportionality of the amount of displacement with current excitation. 

The advantage of the PCA is that despite the similarity of the curves in Fig.~\ref{Specs}{\bf a} on the one hand, as well as the presence of fluctuations for the different measurements, only the changes correlated to the switching events are identified and associated with a small number of principal components. The scores of these components may then serve as a measure for the amount of change related  to the external modification of the system.

\bibliographystyle{naturemag}
\bibliography{RefsBK,Refs}

\newpage

\begin{figure}[h]
  \hspace*{-0cm}\includegraphics[width=\textwidth]{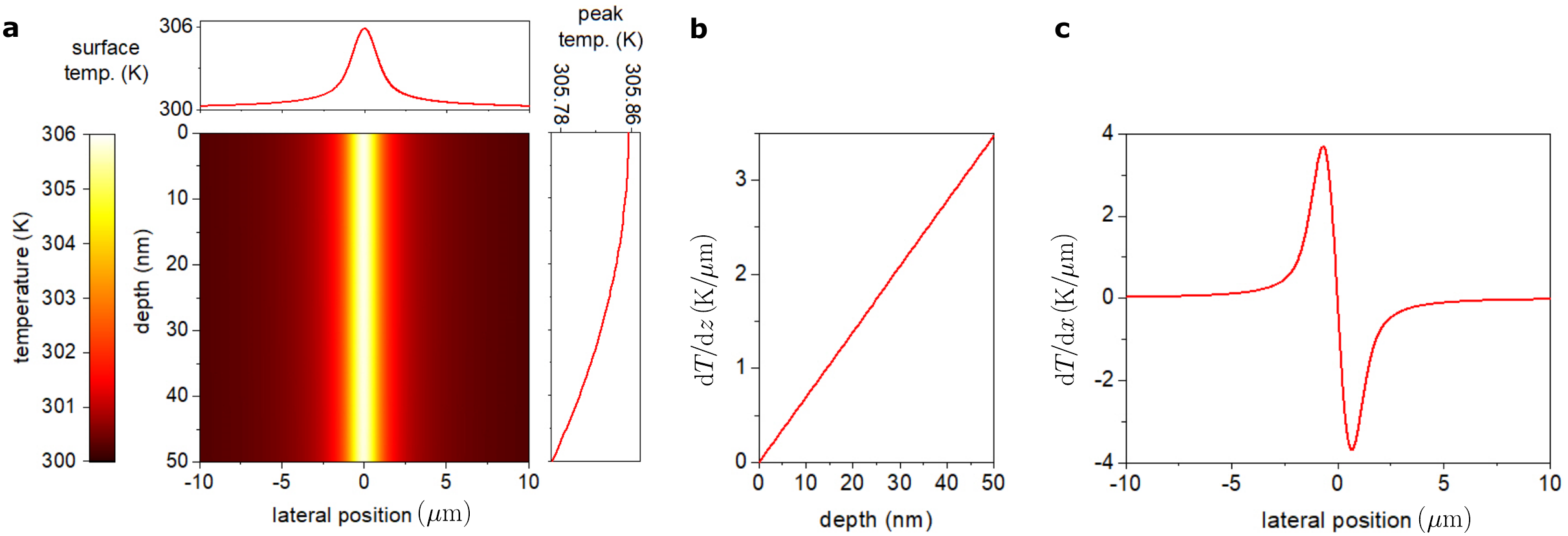}
  \caption{{ \bf Numerically calculated temperature distribution in the CuMnAs sample.}\label{profile} Temperature distribution in a 50~nm thick CuMnAs film grown on top of a 0.5~mm GaP substrate generated by local heating with a focused laser spot of 1.5~$\mu$m FWHM and 5~mW laser power. {\bf a,} in-plane and perpendicular-to-plane temperature profiles.  {\bf b,} temperature gradient normal to the sample plane in the center of the laser spot and {\bf c,} in-plane temperature gradient on the sample surface.
}
\end{figure}

\clearpage
\begin{figure}[h]
  \hspace*{-0cm}\includegraphics[width=.9\textwidth]{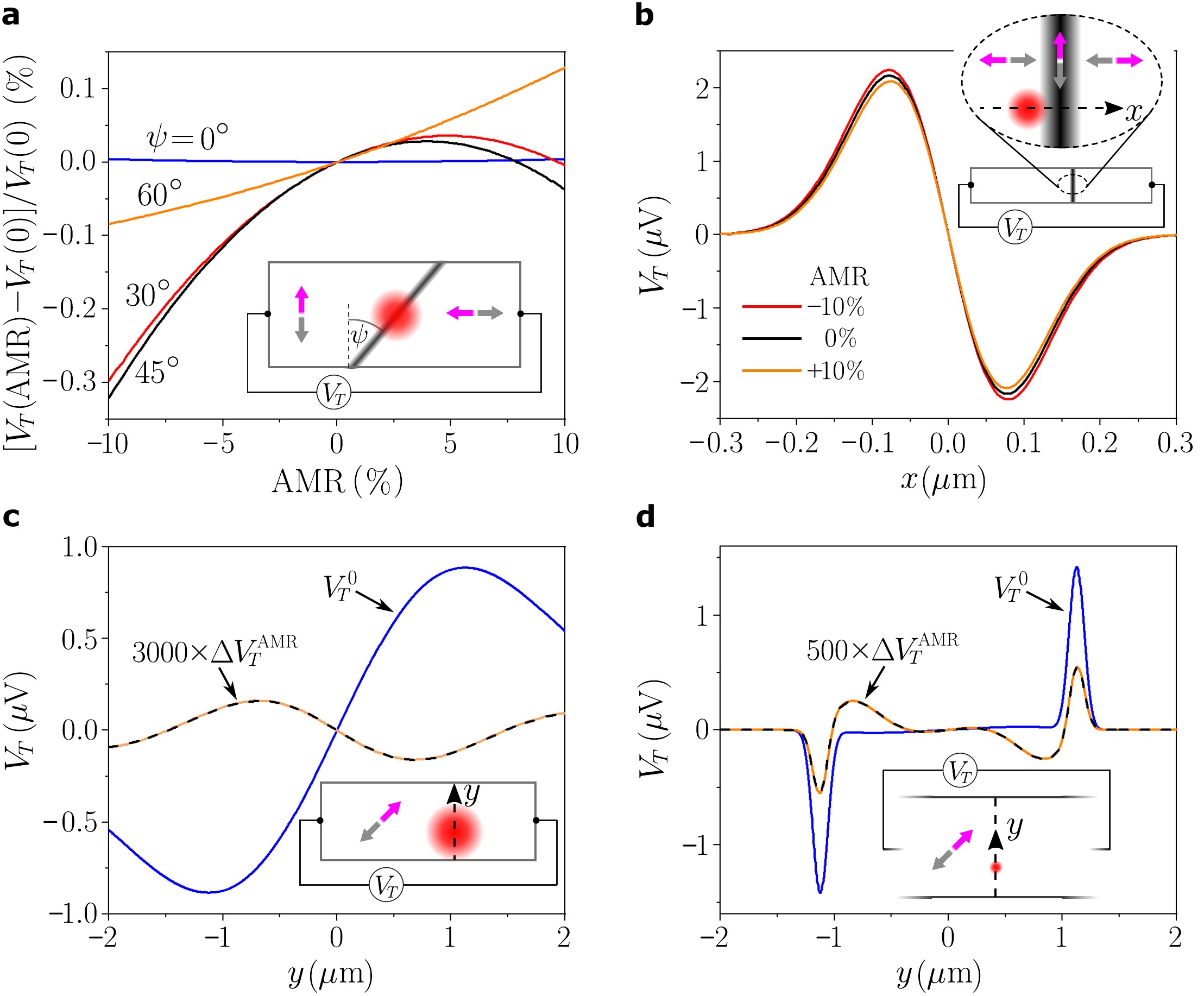}
  \caption{{\bf Effect of the anisotropic magnetoresistance on the thermoelectric voltage generated by the magneto-Seebeck effect.}
  {\bf a,} Dependence of the thermoelectric voltage signal on conductivity variations generated by the $AMR=(\sigma_\parallel-\sigma_\perp) /(\sigma_\parallel+\sigma_\perp)$ plotted relative to $V_T(\mathrm{AMR}\!=\!0)$. $\sigma_{\rm \parallel}$ ($\sigma_p$) is the conductivity in the direction collinear (perpendicular) to the N\'eel vector (indicated by double arrows). The thermoelectric signal is generated when the  laser spot with $1.5\,\mu$m FWHM is focused on the center of a $100\,$nm wide $90^\circ$ domain wall crossing the  $10\,\mu$m wide channel at an angle $\psi$ (see the inset). {\bf b,} Thermoelectric voltage signal with no AMR and $\pm10\,\%$ AMR when a light beam with Gaussian intensity profile of $100\,$nm FWHM is crossing the $100\,$nm wide $180^\circ$ domain wall. {\bf c,} Thermoelectric voltage $V_T^0=V_T(\mathrm{AMR}\!=\!0)$ due to the transverse MSE alone when a laser spot with Gaussian intensity profile of  $1.5\,\mu$m FWHM is crossing a $2\mu$m wide bar in a single antiferromagnetic domain configuration with the N\'eel vector oriented $45^{\circ}$ from the bar axis. Change of the thermal voltage caused by AMR of $\pm10\,\%$ relative to the case of zero AMR, $\Delta V_T^{AMR}=V_T(\mathrm{AMR}\!=\!\pm10\,\%)-V_T(\mathrm{AMR}\!=\!0)$, is shown by the dashed line. {\bf d,} Calculations analogous to {\bf c} for the case of a light spot with Gaussian intensity profile of $100\,$nm FWHM.
\label{anisotropy}}
\end{figure}

\clearpage
\begin{figure}[h]
  \hspace*{-0cm}\includegraphics[width=.9\textwidth]{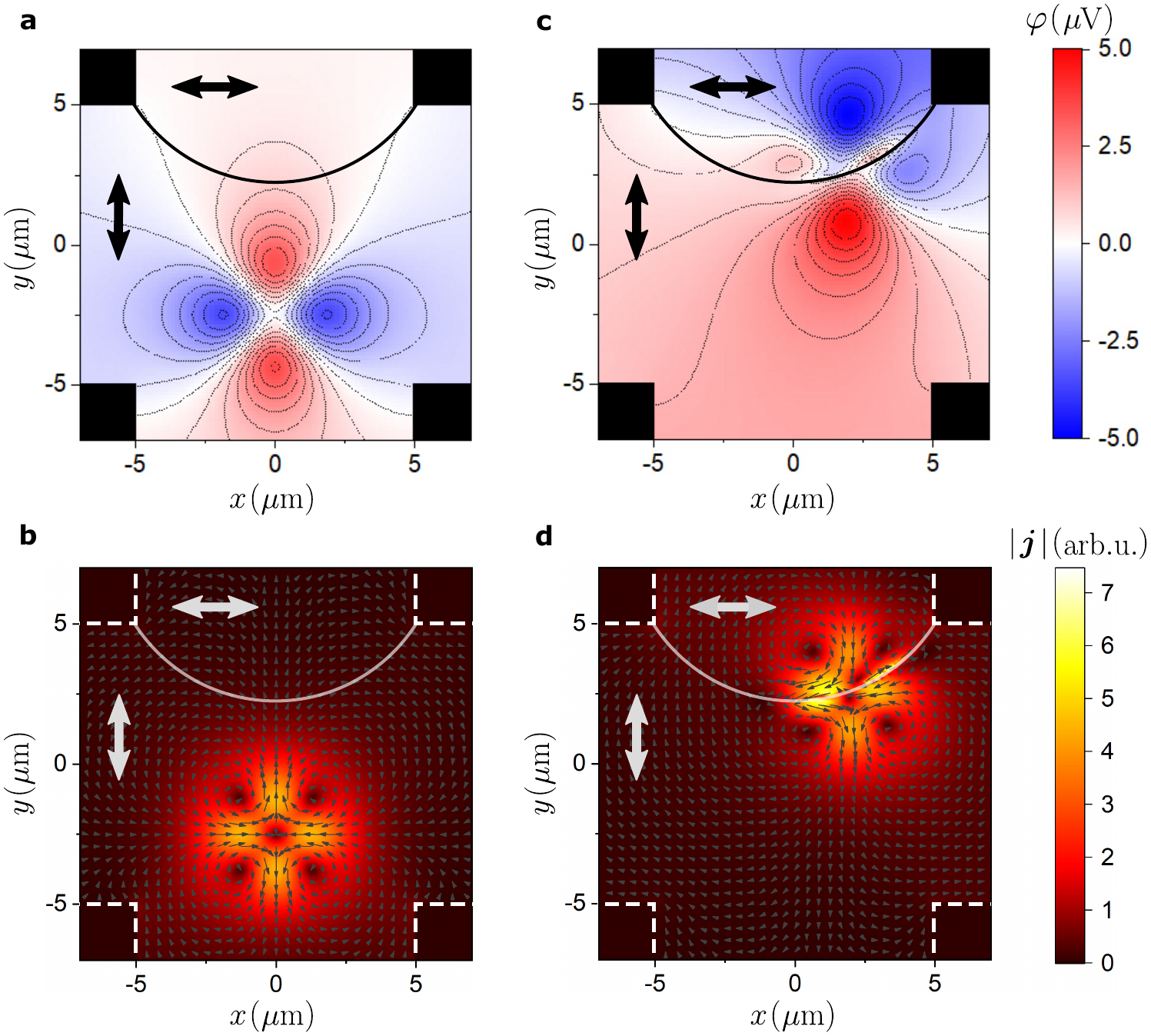}
  \caption{{\bf Electrostatic potential (a,c) and current density (b,d) in a cross-bar structure.}
    Distributions of the electrostatic potential and the current density when the sample is locally heated by a laser beam with a Gaussian intensity profile of 1.5\,$\mu$m FWHM at two different  positions: inside a domain ({\bf a, b}), and,  on top of an arc-shaped domain wall ({\bf c, d}). The domain wall position is marked by a black (white) arc and N\'eel vector orientations are indicated by double arrows. The color scales in {\bf a, c}  and  {\bf b, d}  show the local electrostatic potential  and  the magnitude of the current density, respectively, and the orientation of the local current density is depicted by black arrows.
    \label{potential}}
\end{figure}

\clearpage
\begin{figure}[h]
\hspace*{-0cm}\includegraphics[width=.9\textwidth]{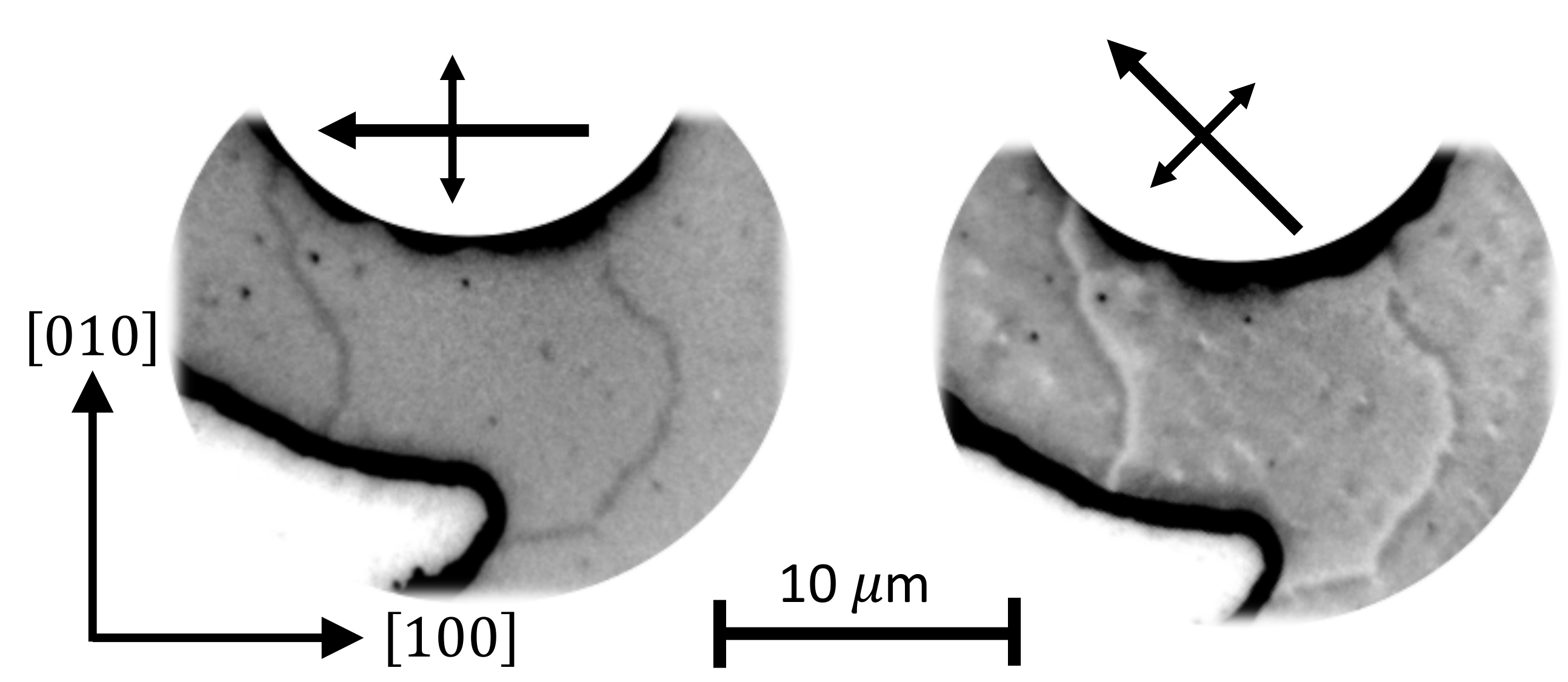}
  \caption{{\bf 180$^\circ$ domain wall in a CuMnAs film with uniaxial anisotropy.}
Antiferromagnetic domain structure of a $10\,$nm thin CuMnAs film, observed with XMLD-PEEM. {\bf left}: The dark lines originate from  180$^\circ$  domain walls, which separate antiferromagnetic domains  of reversed N\'eel vectors visible as extended gray areas. The double arrow indicates the X-ray linear polarization orientation  with respect to the crystallographic orientation of the epitaxial CuMnAs film.  {\bf right}:  Same measurement with $45^\circ$ rotated X-rays polarization. The 180$^\circ$  domain walls appear now as a light-dark contrast.
}
\label{uniaxial}
\end{figure}

\clearpage
\begin{figure}[h]
\hspace*{-0cm}\includegraphics[width=.9\textwidth]{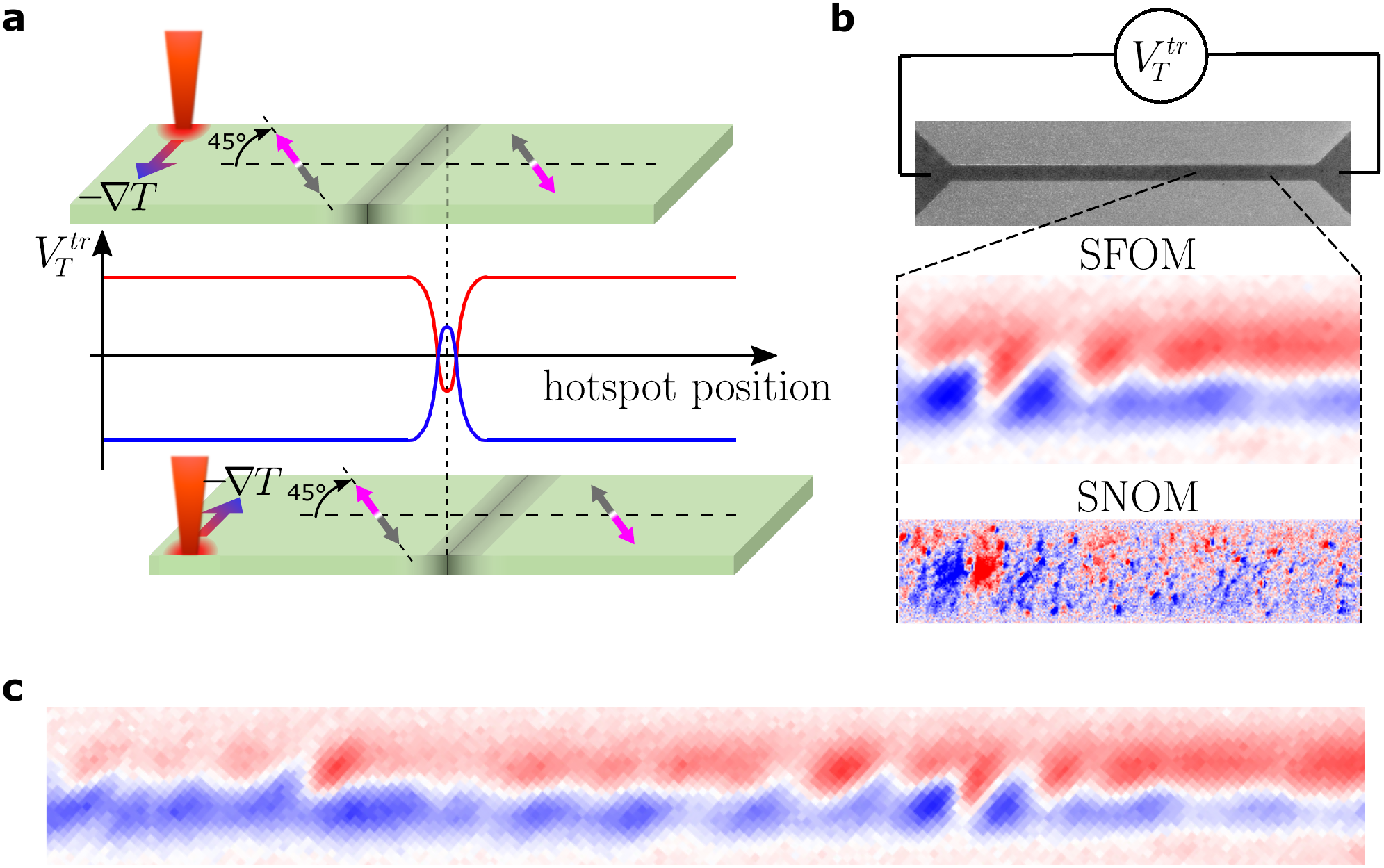}
  \caption{{\bf Transverse magneto-Seebeck effect.} {\bf a,} Schematic of  transerve magneto-Seebeck effect along the edges of a bar device. Assuming a constant N\'eel vector inside the bar device, the two edges show an opposite voltage due to the opposite net temperature gradient. {\bf b,} SFOM- and SNOM-MSE scans of a uniaxial layer where the easy axis lies at $45^\circ$ with respect to the bar direction. {\bf c,} SFOM scan of the entire bar device showing the transverse MSE signals of opposite polarities along the bar boundaries. The additional negative (blue) and positive (red) signal towards the left and right ends of the bar is caused by the ordinary Seebeck effect due to the tiny asymmetry of the heat flow in the horizontal direction from the hotspot towards widening of the bar at left and right contact area (see the SEM micrograph in {\bf b}).
}
\label{transverse}
\end{figure}

\clearpage
\begin{figure}[h]
	\hspace*{-0cm}\includegraphics[width=.9\textwidth]{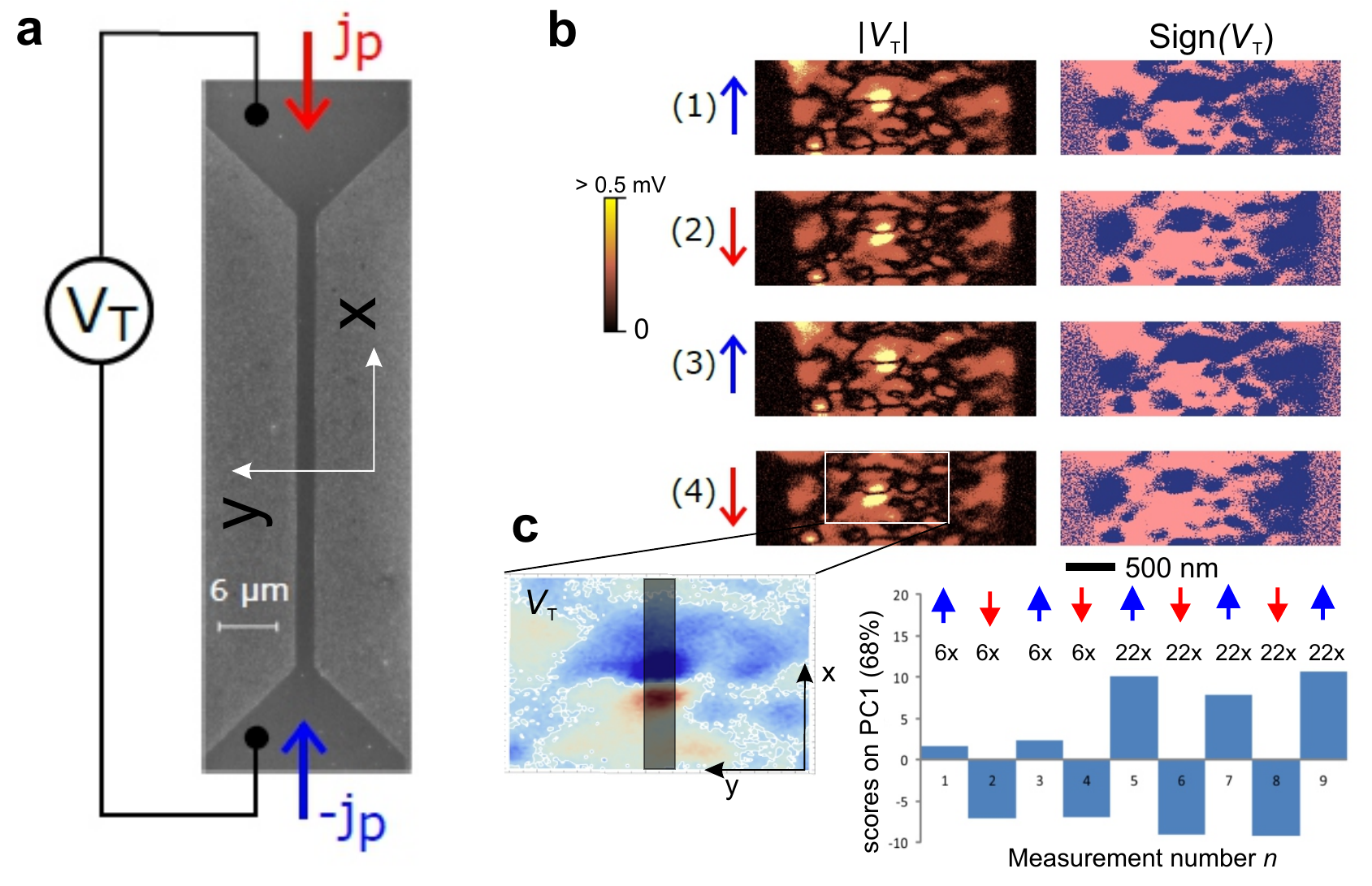}
  \caption{{\bf Switching analyzed by PCA.} {\bf a,} SEM image of the CuMnAs bar device with current-pulse directions indicated by the arrows. {\bf b,} Maps of the thermoelectric voltage $V_T$ for different magnetic states after pulsing with the current polarity indicated on the left. {\bf c,} The section used for PCA with the scores on the principal component PC1 as bar diagram.
	}
	\label{PCA}
\end{figure}

\clearpage
\begin{figure}[h]
	\hspace*{-0cm}\includegraphics[width=.9\textwidth]{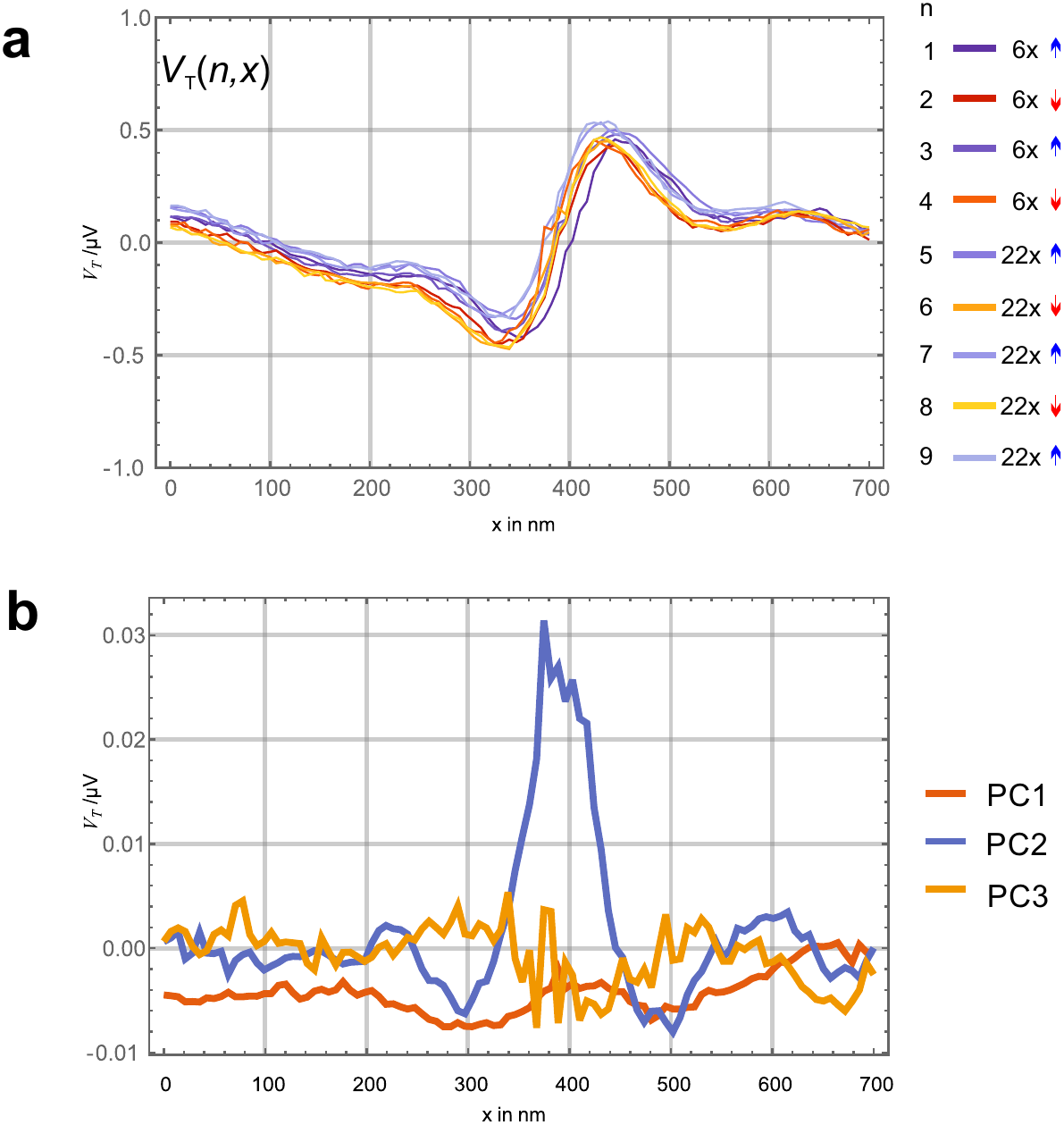}
  \caption{{\bf $\boldsymbol{V_T}$ variation along the axis analyzed by PCA.} {\bf a,} for different magnetic states, {\bf b,} set of basis functions determined by PCA.
	}
	\label{Specs}
\end{figure}